\documentclass[iop]{emulateapj}                

\usepackage{natbib}
\usepackage{threeparttable}
\usepackage{subfigure}
\usepackage{amsbsy,amsmath}
\usepackage{hyperref}
\usepackage{graphicx}
\usepackage{epstopdf}
\usepackage{natbib}
\usepackage{rotating} 

\newcommand{\EQ}{\begin{equation}}
\newcommand{\EN}{\end{equation}}
\newcommand{\EQA}{\begin{eqnarray}}
\newcommand{\ENA}{\end{eqnarray}}

\newcommand{\FF}{\mbox{\boldmath $F$} {}}

\begin{document}
\pagestyle{plain}
\pagenumbering{arabic}

\title{Galaxy Outflows Without Supernovae}
\author{Sharanya Sur\altaffilmark{1,2}, Evan Scannapieco\altaffilmark{2}, Eve C. Ostriker\altaffilmark{3}}
\altaffiltext{1}{Indian Institute of Astrophysics, 2nd Block, Koramangala, Bangalore 560034, INDIA}
\altaffiltext{2}{School of Earth and Space Exploration, Arizona State University, 
PO Box 876004, Tempe - 85287, USA}
\altaffiltext{3}{Department of Astrophysical Sciences, Princeton University, Princeton, NJ 08544, USA}
\email{sharanya.sur@iiap.res.in}

\begin{abstract}

High surface density, rapidly star-forming galaxies are observed to have $\approx 50-100\,{\rm km\,s^{-1}}$ 
line-of-sight velocity dispersions, which are much higher than expected from supernova driving alone, but 
may arise from large-scale gravitational instabilities. Using three-dimensional simulations of local regions of 
the interstellar medium, we explore the impact of high velocity dispersions that arise from these disk instabilities. 
Parametrizing disks by their surface densities and  epicyclic frequencies, we conduct a series of simulations 
that probe a broad range of conditions. Turbulence is driven purely horizontally and on large scales, neglecting 
any energy input from supernovae. We find that such motions lead to strong global outflows in the highly-compact 
disks that were common at high redshifts, but weak or negligible mass loss in the more diffuse disks that are 
prevalent today. Substantial outflows are generated  if the one-dimensional horizontal velocity dispersion exceeds 
$\approx 35\,{\rm km\,s^{-1}},$ as occurs in the dense disks that have star formation rate densities above  
$\approx 0.1\,{\rm M}_\odot\,{\rm yr}^{-1}\,{\rm kpc}^{-2}.$ These outflows are triggered by a thermal 
runaway, arising from the inefficient cooling of  hot material coupled with successive heating from turbulent driving. 
Thus, even in the absence of  stellar feedback, a critical value of the star-formation rate density for outflow generation 
can arise due to a turbulent heating instability. This suggests that in strongly self-gravitating disks, outflows may be 
enhanced by, but need not caused by, energy input from supernovae.

\end{abstract}


\section{Introduction}\label{intro}

Multi-wavelength observations reveal the existence of massive, galaxy-scale outflows of multiphase material, 
driven from rapidly star-forming galaxies  \citep{HAM90, Bomans+97, Pettini+01, Weiner+09, Martin+13, Rubin+14}. 
Such outflows  are thought to play a fundamental role in galaxy evolution: enriching the intergalactic medium (IGM) 
\citep{SC96, SSR02, Pichon+03, Schaye+03, Ferrara+05, Scann+06, Steidel+10, Martin+10}, shaping the galaxy 
mass-metallicity  relation \citep{DS86, Tremonti+04, Erb+06, KE08, Mannucci10}, and affecting the baryonic content 
and number density  of dwarf galaxies \citep{SP99, CLBF00, SFM02, Bens+03}. Outflows also aid in eliminating 
small-scale random magnetic fields from galactic disks \citep{SSSB06, SSS07, CSSS14}, thereby preventing the 
large-scale dynamo from undergoing catastrophic quenching. Yet despite their  importance, understanding the 
dynamical and microphysical processes that control the evolution of galaxy outflows remains a challenge. 

One difficulty in understanding these processes stems from the complex physics of the interstellar medium (ISM),
where heating by UV photons, cosmic rays, and supernova shocks operate  in combination with 
radiative cooling processes, leading to a multiphase, supersonic medium. Given the extremely short cooling times 
and the importance of small physical scales, simulations with $\approx$ parsec resolution are required to model the 
multiphase gas distribution and expansion of supernova remnants directly \citep[e.g.,][]{dAB04, Hill+12, HI14, Walch+14, 
Giri+15, KO15}.  In particular, neither the correct mass of hot gas nor the correct momentum injected to the ISM by 
supernovae can be captured unless both the Sedov and shell formation stages are sufficiently well resolved \citep{KO15}. 
Thus, galaxy-scale numerical simulations have instead relied on a number of approximations, including: temporarily 
lowering the densities and cooling rates of heated particles \citep{GI97, TC00,  Stinson+06, Governato+07}, using an 
empirical heating function \citep{MMN89, MF99}, imposing an artificial  temperature floor \citep{Suchkov+94, TT98, 
Fujita+04}, and implementing exaggerated momentum kicks  \citep{NW93, MH94}, and in the case of feedback from 
active galactic nuclei, storing the energy until it is sufficient to raise the temperature of the surrounding gas above a 
threshold value \citep{VS12}.

The connection to observations is further complicated by recent claims that instead of supernovae, outflows could 
primarily  be driven by radiation pressure on dust \citep{TQM05, MMT11,Hopkins+11,Hopkins+12} or by 
non-thermal pressure contributed by  cosmic rays \citep{Soc+08, SSS10, U+12, Booth+13, Han+13, SB14}. 
In addition, the typically high ISM Reynolds numbers of  ${\rm Re} \approx 10^{5}$ or greater, imply that a complete 
understanding of these massive outflows cannot be obtained by ignoring small-scale turbulent structures. 
Interestingly, using a numerical sub-grid model for the unresolved turbulent velocities and length scales, 
\citet{SB10} showed that it was  possible to produce outflows of multiphase material arising from simultaneous 
turbulent heating and radiative cooling in the disk. Notwithstanding the approximate nature of the scheme 
employed, this offers support to the standard picture that massive, galaxy scale outflows can indeed be 
produced from  turbulence in the disk. 

One hint as to the physics of galaxy outflows may lie in the properties of the galaxies that host them.  
Outflows are observed over a wide range of galaxy masses, but in a smaller range of galaxy surface densities. 
Large outflows are ubiquitous in galaxies in which the star-formation rate density per 
unit area exceeds a critical value of $\dot{\Sigma}_\star^{\rm cr} \approx 0.1\,{\rm M}_\odot\,{\rm yr}^{-1}\,{\rm kpc}^{-2}$ 
\citep{Heckman02, H03}, while the ejection of material is more sporadic for 
$\dot{\Sigma}_\star < \dot{\Sigma}_\star^{\rm cr}$ values \citep{Chen+10}. 
Recent observations also show that disks with strong outflows are characterized by velocity dispersions between
$\sigma_{\rm v}^{\rm 1D} \approx 50-100\,{\rm km\,s^{-1}}$ \citep{Genzel+11,Swinbank+11}. Such high  
velocities are difficult to obtain from supernovae acting alone, as high-resolution simulations of the ISM with a wide 
range of SN rates show that such explosions can only drive velocities to
$\approx 10-20\,{\rm km\,s^{-1}}$ 
\citep{DBB06,JM06,JMB09,KKO11,KOK13, SO12, HI14, Gatto+15, Martizzi+15}. Therefore, how does 
one account for such high velocity dispersions in these disks?

A possible solution lies in the gravitationally-driven motions that occur in high surface density disks. The Toomre 
stability criterion \citep{T64} relates the total disk surface density $\Sigma$, the epicyclic frequency $\kappa$, and 
the sound speed $c_s$ in infinitesimally thin disks that are marginally unstable to axisymmetric modes as
$Q \equiv \kappa\,c_{s}/\pi\,G\,\Sigma = 1$, where $G$ is the gravitational constant. Allowing for 
non-axisymmetric instabilities, magnetic fields, and interaction with a stellar disk increases the critical $Q$ value, 
while thick disk effects decrease it \citep[e.g.,][]{Romeo92, KO01, KOS02, KO07, RF13}. 
In the real ISM, turbulent velocities are comparable to the thermal sound speed of the 
warm medium and are greater than the thermal sound speed of the cold medium. Thus any characterization of the 
effective $Q$ should depend on the total (thermal plus scale-dependent turbulent) velocity dispersion \citep{RBA10,HR12}.
In addition, ISM turbulence is driven by a combination of feedback from star formation and gravitational instabilities, 
and characterizing stability in realistic disks is therefore quite complex \citep{ARG15}. However, both Milky-Way 
type galaxies simulations \citep{WMN02, Agertz+09, ARG15} and high-redshift galaxy simulations
\citep[e.g.,][]{Imm+04, CDB10, Genel+12, GDC12} show that at sufficiently large scales 
gravitational instabilities promote an increase of the velocity dispersion, $\sigma,$ until 
\EQ
Q_{\rm eff} \equiv [\sigma^{2} + c_{s}^{2}]^{1/2} \kappa/\pi G\Sigma \approx 1.
\EN 
This implies that disks with high surface densities {\it must} develop significant turbulent motions even if stars are 
unable to stir the disks sufficiently to stabilize them, because gravitational instabilities will lead to the formation of 
clumps moving at typical velocities $\sigma \approx \pi\,G\,\Sigma/\kappa.$

Based on these ideas, \citet{SGP12} conducted simulations of turbulently stirred, radiatively cooled media. These 
simulations modeled a local patch of the galaxy as a stratified medium in which turbulence was  driven at a rate 
that matched the overall cooling rate.  At low velocity dispersions, such as occur in the Milky Way, this configuration 
was stable for many dynamical times. On the other hand, the critical star-formation rate density for galaxy outflows 
corresponds to a gas surface density of $\Sigma_{\rm g} \approx 100\,{\rm M}_\odot\,{\rm pc}^{-2}$, which assuming 
typical values of $\Sigma \approx 2\,\Sigma_{\rm g},$ $c_s\approx10\,{\rm km\,s^{-1}}, $ and $\kappa^{-1}\approx 15\,{\rm Myr}$ 
gives one-dimensional turbulent velocity dispersion of $\sigma \approx 35\,{\rm km\,s^{-1}}$ to have $Q_{\rm eff} \approx 1$.
At these high dispersions, \citet{SGP12} discovered  the onset of a thermal runaway, where multiphase material moved 
upward from the disk and out of  the simulation domain, implying the absence of a stable equilibrium beyond this critical 
value. 

In this paper, we examine effects of turbulence in Toomre-critical disks more closely by conducting three-dimensional 
numerical simulations of a local patch of the ISM for a range of values of the total surface density $\Sigma$  and the 
epicyclic frequency, $\kappa$. Our goal is to better understand the varying environments in galaxies of different surface 
densities, and the extent to which these variations lead to the direct driving of  galaxy-scale outflows and to conditions 
that are favorable to the driving of outflows by supernovae. Motivated by the idea that large-scale gravitational instabilities 
maintain a level of turbulence in which $Q_{\rm eff}\approx 1$, 
and that these instabilities primarily involve in-plane motions, we apply driven turbulence at a forcing level that 
results in a horizontal velocity dispersion
\EQ
\sigma_{\rm H}\approx \pi\,G\,\Sigma/\kappa.  
\label{horzvel}
\EN 
As we shall show, the horizontally-driven turbulence also leads to vertical motions (at a lower amplitude), and the shocks 
from both horizontal and vertical motions heat the gas. Thermal pressure gradients and vertical turbulence combine to 
drive outflows from the disk. To focus on the driving of turbulence purely by gravitational instabilities, we do not 
include any vertical mechanical or thermal energy input from supernovae in this study, such that all outflows obtained 
represent a lower limit over which stellar processes will lead to additional contributions.

Because our study is focused on obtaining a better understanding of the role of gravitational instabilities in 
changing the nature of the medium in which stellar processes operate, we deliberately do not attempt to re-create 
a full model of the ISM, in which feedback from supernovae \citep{DBB06,KKO11,SO12}, ionization fronts
\citep{Matz02,Walch+12}, chemical transitions \citep{KI00,Walch+11,dAB12}, radiation pressure 
\citep{KT12,Sales+14}, cosmic rays \citep{Zirak+96, Boett+13, Booth+13} and magnetic fields 
\citep{Gressel+08, Gent+13} all play a role.  Although simulations that include this physics would be closer to 
real galaxies, they would also be much more difficult to interpret, as we could never be sure 
how to connect causes and effects unambiguously. 
In this sense, simulations that start from simpler initial conditions 
and explore the role of a few free parameters are complementary to more complex ISM simulations 
in the science they are able to target.

The structure of this paper is as follows.  In Section 2 we discuss our numerical methodology. In Section
3 we describe our results, focusing on outflow rates and phase distribution of the gas, turbulent properties, and 
the cooling and free-fall times in the media. Conclusions are  presented in Section~4.


\section{Numerical Modeling}\label{num}

Our simulations contain only four components: (i) the equations of 
compressible fluid dynamics; (ii) a continuously updated average  vertical gravitational acceleration 
 to capture the evolution of the disk scale height in response to thermal  pressure and vertical turbulence; 
(iii) radiative cooling of atomic gas in the optically-thin limit; and (iv) purely horizontally-driven turbulence 
that approximates the impact of gravitational instabilities in a rotating disk in the absence of stellar feedback. 

A self-gravitating disk with velocity dispersion $\sigma$ has a scale height
$H=\sigma^2/(\pi G \Sigma)$. Taking this as a characteristic turbulent 
forcing scale $\lambda_{\rm f}$ and assuming that the effective Toomre parameter is unity so that 
eq.\ (\ref{horzvel}) holds, we have
\EQ
\lambda_{\rm f} \approx R \equiv \pi\,G\,\Sigma/\kappa^{2},
\label{stirscale}
\EN 
for the turbulent {\it stirring} scale. In reality, power may be driven by gravitational instabilities over 
a range of scales and will cascade to smaller scales via nonlinear interactions.  Here, we chose 
a characteristic spatial scale for simplicity.  We note that the adopted stirring scale is comparable 
to the range where the turbulent power spectrum reaches its maximum in the shearing-box simulations 
of \citet{KO07}, who found that the power is flat above $\lambda_x \approx 8 G \Sigma/\kappa^2$.

The simulations were conducted with the multidimensional, grid-based (magneto)-hydrodynamic 
code FLASH  (version 4.2) \citep{Fryxell+00}. While FLASH is capable of incorporating dynamical grids of 
varying resolution by virtue of the adaptive mesh-refinement (AMR) technique, we chose to perform our 
simulations  on a uniform grid with the unsplit hydrodynamic solver \citep{Lee+09,Lee+12}, in a box of size 
$3R$ in the $x$ and $y$ directions and $-3R$ to $+3R$ in the vertical direction, where $R$ is defined by 
eq.\ (\ref{stirscale}). 
We adopted a $256^{2} \times 512$ grid for the majority of our simulations, and, for the purpose 
of a resolution study,  we also conducted simulations at resolutions of $64^{2} \times 128$, $128^{2} \times 256$ 
and $512^{2} \times 1024$ (see Appendix). Furthermore, we chose periodic boundary conditions in $x$ and $y$ 
and `diode' boundary conditions in the $z$ direction, which allow for outflows of material from the simulation 
domain, but prevent inflows. 

The initial conditions were characterized by two free parameters: the total matter  surface density, $\Sigma$ 
and the epicyclic frequency, $\kappa$. We further assumed that the gas surface density was given by
$\Sigma_{\rm g} = f_{\rm g}\,\Sigma$, where  the gas fraction, $f_{\rm g},$  was fixed at 1/2 for all our 
simulations. Because we do not know {\em a priori} what vertical distribution the medium will take at late 
times, we adopted an initial density distribution that maintained the desired surface density and 
approximated the vertical distribution as an exponential profile:
\EQ
\rho = \frac{\Sigma}{2 R/C} \exp\left( \frac{-|z|}{R/C}\right) \frac{1}{1- \exp(-3C)},
\label{eq:densityprofile}
\EN 
where $C$ is a `compression' factor that relates $R$ and the initial scale height such that $H(t=0) =R/C,$
and the  $1- \exp(-3C)$ insures that the total surface density within the finite domain is equal to $\Sigma.$
In the gradual course of the simulation, and in the absence of any other feedback processes, the scale height, 
$H$, the vertical velocity dispersion, $\sigma_{\rm z}$, and the disk thermal structure adjusted automatically to 
the most compact and coldest distribution available for a given choice of $\Sigma$ and $\kappa$. If the 
turbulent motions were isotropic and thermal pressure support were minimal, then we would expect our 
simulations to reach a state-state density distribution similar to eq.\ (\ref{eq:densityprofile}) with $C \approx 1.$ 
However, if vertical turbulent motions remain smaller than horizontal ones, as we shall see is the case, then 
we would expect much more compact distributions, corresponding to eq. (\ref{eq:densityprofile}) with larger 
$C$ values. For the simulations presented in this paper, we have used $C = (5 - 7)$ at $t=0$. In general, this 
is somewhat more compact than the final steady-state distributions described below, 
but because our gas distribution collapses while the turbulence is developing, we find that smaller choices 
of $C$ lead to longer delays in reaching the same  quasi-steady state. In addition, in all our simulations, we 
employed an ideal gas equation of state with $\gamma = 5/3$ and an atomic mass of one, an initial constant 
temperature of $T= 2\times10^{5}\,{\rm K},$ and zero initial velocities.
 
Because our simulations were not conducted in a shearing-box in which large scale gravitational 
forces are opposed by centrifugal and Coriolis forces, and because we do not incorporate sink particles to 
handle the collapse of Jeans-unstable gas, we cannot simply implement self-gravity in our simulations.
Thus we adopted an idealized approach. First, to include the impact of gravity on the overall vertical profile of 
the gas, we computed a vertical acceleration that was a time-dependent function of $z$, and did not depend on 
$x$ and $y$. The time dependence ensures that the system finds its own vertical scale-height in course of the 
simulation, rather than having the user guess a scale-height a priori, as would be the case if the user were to 
specify a gravitational profile at runtime. Second, we added a horizontal stochastic driving term, to include the 
impact of large-scale horizontal turbulent motions driven by gravity. 

To compute the vertical gravity, at every time step we calculated a vertical acceleration 
profile, $g(z),$ directly from the average vertical mass profile as
\EQ
g(|z|) = 2\pi G\int_{-|z|}^{|z|}\left|\frac{d\Sigma}{dz}\right | dz,
\EN 
where we approximate $d\Sigma/dz$ as $(1/f)d\Sigma_{\rm g}/dz$. 
Because this acceleration is a function of height and not of horizontal position, it does not directly drive turbulence.
To implement turbulent driving as would be induced by large-scale gravitational instability,
we used a version of the stirring module in FLASH, adapted from 
one of our recent studies \citep{Sur+14}, which models  turbulent driving random motions as a stochastic 
Ornstein-Uhlenbeck (OU) process  \citep{EP88, Benzi+08}. Specifically, this corresponds to a Gaussian random 
vector field, $\FF$, with an exponential  temporal correlation, $t_{\rm f}$, in the momentum equation. 

We drove turbulence  in the range of wavenumbers $2\leq |{\bf{k}}|\,L/2\pi \leq 3$, such that the average forcing 
wavenumber was $k_{\rm f}\,L/2\pi\simeq2.5$.  Here $L = 3R$ is the horizontal extent of the box and the wave 
vector $|{\bf k}| = \sqrt{(k_{x}^2 + k_{y}^2 + k_{z}^2)}$. This corresponds to a turbulent driving scale, 
\EQ
L_{\rm f} = 2\,\pi/k_{\rm f} = 1.2\,R. 
\label{fscale}
\EN
Since we wish to restrict driving of turbulence to a two-dimensional plane (to simulate the effect of self-gravity within a disk), we modified our routines to force only the 
horizontal component of the velocity.
The correlation time  of turbulence, $t_{\rm f},$ was chosen to be  $\approx 0.1\,t_{\rm ed}$, where 
\EQ
t_{\rm ed} = L_{\rm f}/\sigma_{\rm H} = 1.2\,\kappa^{-1},
\label{teddy}
\EN 
is the eddy-turnover time. For a given value of $\Sigma$ and $\kappa$, 
the amplitude of the forcing was adjusted  to get the desired value of $\sigma_{\rm H}=\pi\,G\,\Sigma/\kappa$. 
Note that choosing a different value  of $t_{\rm f}$ would require re-adjusting the forcing amplitude 
so as to obtain the correct value of $\sigma_{\rm H}$. 

\begin{figure}[t!]
\centering
\includegraphics[width=1.0\columnwidth]{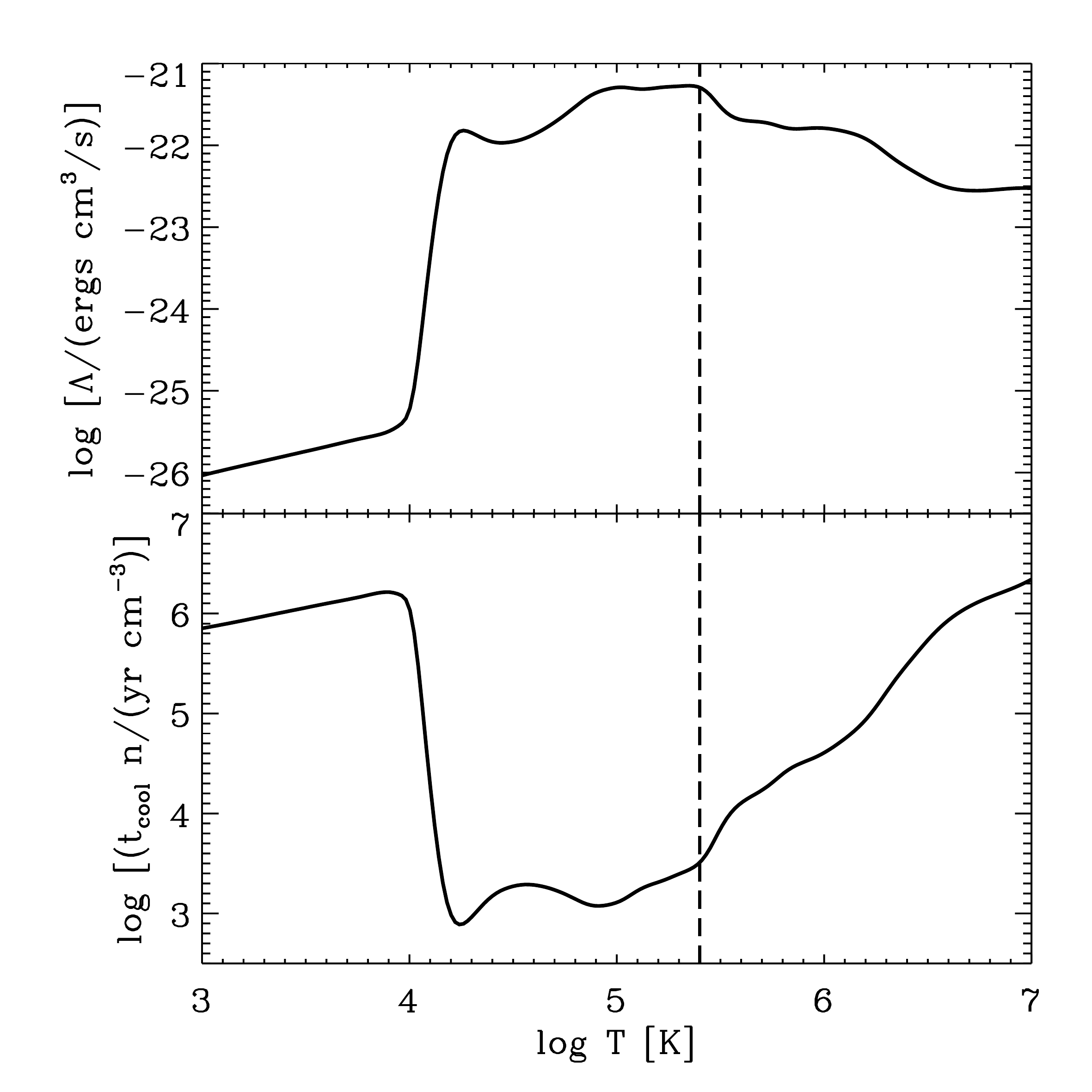}
\caption{Total   atomic, ionic, and Bremsstrahlung
cooling rate (top panel) and the cooling time (bottom panel) as a function of the temperature for solar 
metallicity material. Note that the cooling rate drops above $T\approx 2.5\times 10^{5}\,{\rm K}$, leading 
to an increase in the cooling timescale.}
\label{cool_curve}
\end{figure}

The majority of the energy input from turbulent driving in our simulations leaves the system through radiative cooling, 
as it does in a real galaxy. Following the prescription of \citet{GS10}, we implemented  
atomic, ionic, and Bremsstrahlung cooling
in the optically thin limit, assuming local collisional equilibrium with, 
\EQ
\dot E_{\rm cool}  =  (1-Y)(1-Y/2) \frac{\rho \Lambda(T,Z)}{(\mu m_p)^2}. 
\label{eq:ecool}
\EN
Here $\dot E_{\rm cool}$ is the radiated energy per unit mass, $\rho$ is the density in the cell, $m_p$ is the 
proton mass, $Y$ is the helium mass fraction, $\mu$ the mean atomic mass, and $\Lambda(T,Z)$ is the cooling 
rate as a function of temperature and metallicity.  For the cooling source terms, we implemented a sub-cycling 
scheme \citep{GS10}, such that $T$ and $\Lambda(T,Z)$  were recalculated every time 
$E_{\rm cool}/E > 0.1$. This is equivalent to an integral formalism that assumes a constant density over each 
hydrodynamic time step \citep{TC92, STD01}. Furthermore, in order to help the simulations reach quasi steady-state 
conditions more quickly, cooling is disabled for the first $0.5\,R/\sigma_{\rm H}$ time of the simulation, to avoid 
excessive vertical collapse while the disk gains an initial level of turbulence. 

The cooling rates were calculated using the tables compiled by \citet{WSS09} making the simplifying 
approximation that the metallicity is always solar and that the abundance ratios of the metals 
always occurs in solar proportions.
In the upper panel of Figure~\ref{cool_curve}, we show the total radiative cooling 
rate as a function of the temperature. The lower panel shows the behavior of the local 
cooling timescale defined as the ratio of the energy per unit volume to the radiative cooling rate per 
unit volume, 
\EQ
t_{\rm cool} \equiv 1.5 nk_{B}T/(\rho\,\dot{E}_{\rm cool}) \propto \rho^{-1}, 
\label{tcool}
\EN
where $k_{B}$ is the Boltzmann constant and $n = \rho/\mu\,m_{p}$ is the number density of the gas.

Following the nomenclature introduced by \citet{Wolf+95, Wolf+03}, much of the interstellar medium can be described 
as comprised of a cold neutral medium (CNM) with $T  \approx 100{\rm K}$, a warm neutral medium (WNM) with 
temperatures in the range, $T \approx 6\times 10^{3} - 10^{4}\,{\rm K}$ and a hot ionized medium (HIM) with 
$T \geq 10^{6}\,{\rm K}$. As Figure~\ref{cool_curve} shows, we have only considered cooling in the temperature 
range $T \approx 10^{3} - 10^{7}\,{\rm K}$ and up to a maximum density of $n_{\rm max} = 6.0 \times 10^{7}\,{\rm cm^{-3}}$.
We note that the colder parts of the ISM occupy a small fraction of the volume (only a few percent), and while they 
are important for star formation, they interact much less strongly with the hot medium than the warm gas that occupies
most of the volume. 
Since our goal is to capture the formation of galaxy outflows which are mainly comprised of hot gas in the temperature 
range $T \approx 10^{6} - 10^{7}\,{\rm K}$, we to lowest order neglect the inclusion of the cold dense phases of the 
ISM in our simulations. We therefore do not include low-temperature cooling instead placing a temperature floor at 
1000 K throughout the simulations. This therefore amounts to adopting a simplified description of the multi-phase ISM, 
which as we shall show in later sections is sufficient for the physical problem at hand. In a test case, we 
lowered the temperature floor to $300$ K and found the mass outflow rates to remain unchanged (see Appendix). 

At temperatures $T \approx 10^{4}-10^{6.5}\,{\rm K}$, cooling results mainly from  line emission, 
while cooling due to bremsstrahlung (free-free emission) becomes important at temperatures 
$\geqslant 10^{7}\,{\rm K}$. 
Below $10^{4}\,{\rm K}$, collisions are not energetic enough to excite atomic transitions, 
leading to a drop in the cooling rate. 
The atomic cooling rate attains a  peak value of $\Lambda\approx 10^{-21}\,{\rm ergs\,cm^{3}\,s^{-1}}$ in the 
temperature range $T \approx (1 - 2.5)\times10^{5}\,{\rm K}$. In this regime, the cooling time is 
roughly constant as seen in the lower panel of the plot. The peak at $T\approx 2.5\times10^{5}\,{\rm K}$ is 
dominated by line emission from metal ions, whose atomic energy levels are easily excited by collisions 
at this temperature. However, beyond $T\approx 2.5\times10^{5}\,{\rm K}$, most of the atoms become fully 
ionized, the effectiveness of the line cooling decreases, and the cooling rate drops, leading to a 
gradual increase in the cooling time. What this implies for the nature of the multiphase medium will be 
discussed in the next section. 


\section{Results}\label{results}

\subsection{Outflow Rates}

As our goal is to obtain a better understanding of the turbulent ISM as a function of galaxy properties, we conducted
 a suite of simulations with different values of the gas surface 
density, $\Sigma_{\rm g},$ and the epicyclic frequency, $\kappa$. In Table~\ref{sumsim}, we present a summary of 
the properties of these simulations, including the run parameters ($\Sigma_{\rm g}$, $\kappa$),
the key quantities for each model that can be directly derived from these parameters - the forcing scale $L_{\rm f}$, 
box size above the midplane, $z_{\rm max} = 3R$, the resolution $dz= z_{\rm max} /256$, 
the horizontal velocity dispersion $\sigma_{\rm H}$, and the escape velocity $v_{\rm es}$,
and the quantities that can only be measured from 
the full simulations - the vertical velocity dispersion $\sigma_{\rm z}$, the scale-height $H$, and the mass loss 
rate $\dot{\Sigma}_{\rm g}$.
The idealized nature of our simulations  enables us to probe a wide parameter space, with 
$\Sigma_{\rm g} \in [50, 500]\,{\rm M_\odot\,pc^{-2}}$ and  $\kappa^{-1} \in [6.5, 30]\,{\rm Myr}.$ 
Note however that for runs with $\kappa^{-1}=30\,{\rm Myr}$, simulations with an initial 
$\Sigma_{\rm g} > 75\,{\rm M}_\odot\,{\rm pc}^{-2}$ result in box sizes that are larger than the typical size of entire
disk galaxies, and we have therefore chosen to omit such simulations from our study.

\begin{table*}[ht!]
\begin{minipage}{170mm}
\begin{center}
\resizebox{\textwidth}{!}{
\begin{tabular}{|c|c|c|c|c|c|c|c|c|c|c|} \hline \hline 
Simulation & $\Sigma_{\rm g}$ & $\kappa^{-1}$ & $L_{\rm f}$ & $z_{\rm max}$ & $dz$ & $H$ & 
$\bar{\sigma}_{\rm H}^{\rm 1D}$ & $\bar{\sigma}_{\rm z}$ & $v_{\rm es}$ & $\dot{\Sigma_{\rm g}}$ \\
Name & $[{\rm M}_{\odot}\,{\rm pc^{-2}}]$ & $[\rm {Myr}]$ & [pc] & [kpc] & [pc] & 
[pc] & $[\rm km\,s^{-1}]$ & $[\rm {km\,s^{-1}}]$ & $[{\rm km\,s^{-1}}]$ & $[{\rm M}_\odot\,{\rm yr}^{-1}\,{\rm kpc}^{-2}] $ \\ \hline \hline
S500K6.5* & $500$ & $6.5$ &  $723$ & $1.80$ & $7.03$ & $7.0$ & $68$ & $5.0$ & $309$ & $0.22$ \\ \hline
S250K6.5* & $250$ & $6.5$ & $362$ & $0.9$ & $3.50$ & $9.4$ & $35$ & $5.5$ & $155$ & $0.073$ \\ \hline
S150K6.5* & $150$ & $6.5$ & $217$ & $0.53$ & $2.10$ & $17$ & $21$ & $6.3$ & $92$ & $0.04$ \\ \hline
S250K10* & $250$ & $10$ & $856$ & $2.13$ & $8.32$ & $11$ & $50$ & $5.5$ & $238$ & $0.06$ \\ \hline
S150K10* & $150$ & $10$ & $514$ & $1.27$ & $4.96$ & $20$ & $32$ & $6.0$ & $142$ & $0.03$ \\ \hline
S100K10 & $100$ & $10$ &  $342$ & $0.85$ & $3.32$ & $27$ & $21$ & $6.2$ & $95$ & $0.017$ \\ \hline
S150K20* & $150$ & $20$ & $2054$ & $5.12$ & $20.0$ & $22$ & $60$ & $5.0$ & $285$ & $0.03$ \\ \hline
S100K20* & $100$ & $20$ &  $1369$ & $3.40$ & $13.2$ & $26$ & $53$ & $5.0$ & $190$ & $0.02$ \\ \hline
S75K20 & $75$ & $20$ & $1027$ & $2.56$ & $10.0$ & $35$ & $30$ & $6.0$ & $143$ & $0.009$ \\ \hline
S50K20 & $50$ & $20$ & $685$  & $1.70$ & $6.64$ & $53$ & $21$ & $6.3$ & $95$ & $0.004$ \\ \hline
S75K30 & $75$ & $30$ &  $2311$ & $5.74$ & $22.4$ & $33$ & $45$ & $5.8$ & $214$ & 0.005 \\ \hline
S50K30 & $50$ & $30$ &  $1541$ & $3.83$ & $15.0$ & $50$ & $30$ & $6.0$ & $143$ & $0.004$ \\ \hline \hline 
\end{tabular}
}
\end{center}
\caption{  
Summary of the simulation runs at a uniform grid resolution of 
$256^{2} \times 512$. The different columns from the left to the right are - 1) Simulation name, 2) Gas surface density, 
3) Inverse of the epicyclic frequency, 4) the horizontal forcing scale, $L_{\rm f} = 1.2R$, 5) box size above 
the midplane, 6) the resolution, 7) vertical scale height obtained from equation~(\ref{vertscale}), 8) time-averaged 
mass-weighted 1D horizontal velocity dispersion, 9) time-averaged mass-weighted vertical velocity dispersion, 
10) escape velocity of the gas as defined by equation~(\ref{ves}), and 11) the mass loss rate. Simulations that
show persistent outflows for many eddy-turnover times are marked by an *. }
\label{sumsim}
\end{minipage}
\end{table*}

Our motivation for using a broad range of values is to capture the properties of high surface density galaxies over 
a broad range of masses and redshifts.
Several attempts have been made to make  predictions about the evolution of 
the galaxy size with redshift \citep[see][for recent reviews]{Shapley11, SM12, Con14}, and 
observational studies  have shown that distant galaxies are 
more compact than those of the same mass in the nearby Universe 
\citep[e.g.,][]{Daddi+05, Trujillo+07, Buitrago+08, vanDok+10}. From these studies, the dependence of  
galaxy size with redshift selected at a constant stellar mass can be roughly characterized as a power 
law of the form $R \propto (1+z)^{-\alpha},$ although \citet{Ferg+04} find $ R \propto H^{-1}(z)$ from $z \approx 1-5$,  
where $H(z)$ is the Hubble parameter. 
Assuming, for simplicity, a power law relation between the galaxy size and redshift,   it can be 
shown that the epicyclic frequency for a fixed mass, scales as $\kappa\propto \Sigma_{\rm g}^{1/2}\,(1+z)^{\alpha/2},$ 
while the surface density at fixed mass scales as $\Sigma_{\rm g}\propto(1+z)^{2\alpha}$. 
These scalings imply that star-forming galaxies at higher redshifts are both more compact 
(i.e., have larger values of $\Sigma_{\rm g}$) and are more rapidly rotating (i.e., have lower values of $\kappa^{-1}$) 
than their local counterparts.  As we show below, this allows us to make some quantitative statements 
about the role of gravity-driven turbulence in powering outflows in distant galaxies. This is all the more important as 
the integrated star-formation rate peaks at $z\approx 2$ \citep[see][for a review]{MD14}. 

\begin{figure}[ht!]
\centering
\includegraphics[width=1.0\columnwidth]{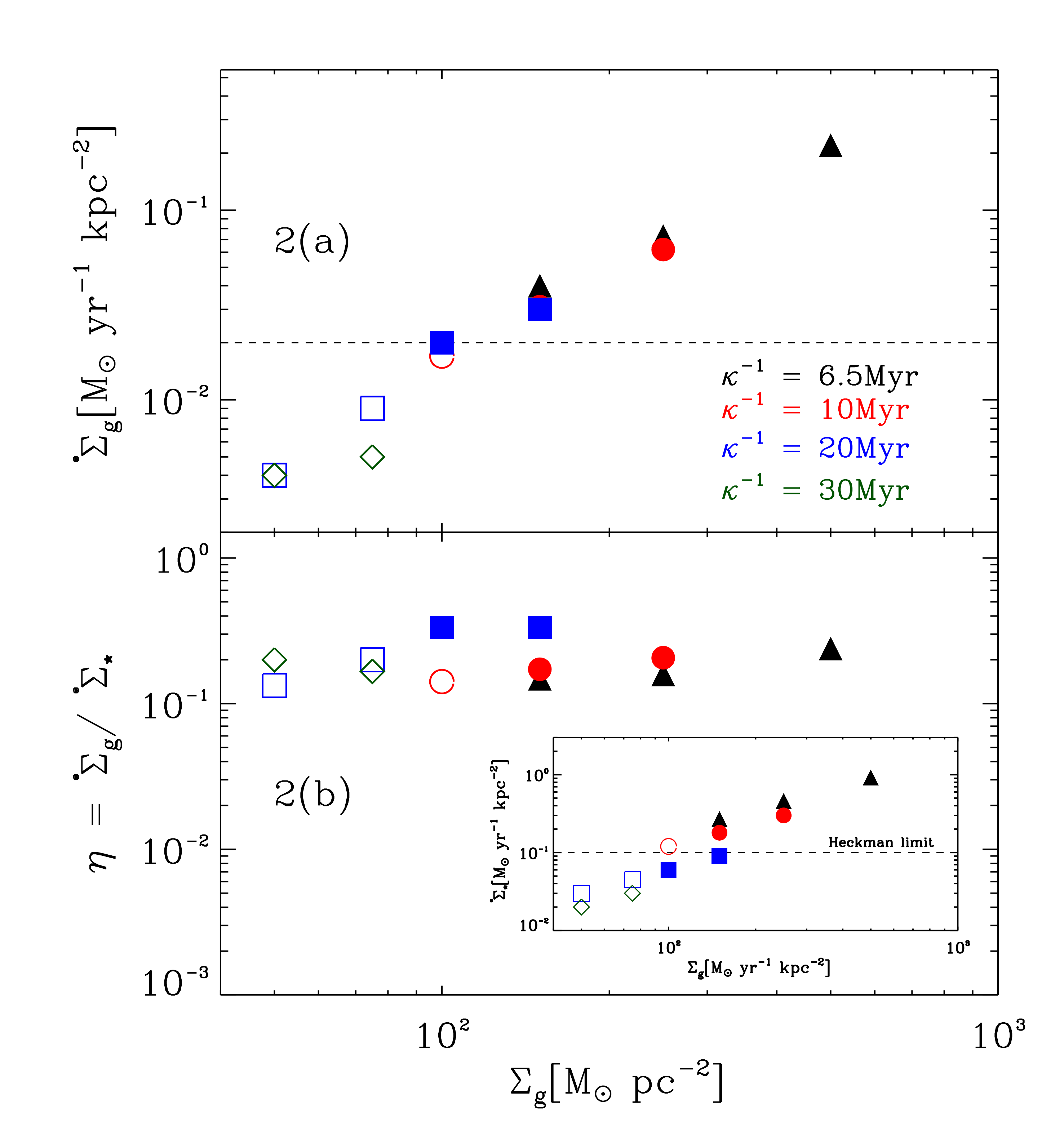} 
\caption{(color online). The upper panel shows the averaged gas mass loss rate as a function of
$\Sigma_{\rm g}$ for different values of $\kappa^{-1}$. Filled symbols denote runs where the computed 
gas mass loss rate $\geqslant 0.02 \,{\rm M_\odot\,yr^{-1}\,kpc^{-2}}$ (horizontal dashed line). The lower 
panel shows the ratio $\eta$ of the gas mass loss rate and the star formation rate (where the star formation 
rate is obtained from equation~\ref{kenlaw}). The inset panel shows the adopted variation of the star 
formation rate as a function of the gas surface density while the dashed line denotes the `Heckman limit'. 
The filled (open) symbols in these panels follow the convention of those in panel 2(a). }
\label{starform}
\end{figure}

In the top panel of Figure~\ref{starform}, we show the variation of the averaged gas mass loss rate as a function of 
$\Sigma_{\rm g}$ for different values of $\kappa^{-1}$. The filled symbols denote runs where we find outflows with
$\dot{\Sigma}_{\rm g} \geqslant 0.02\,{\rm M}_\odot\,{\rm yr}^{-1}\,{\rm kpc}^{-2},$ persisting over many 
eddy-turnover times in disks. Below this threshold, outflows appear to be sporadic or absent with negligible values of 
$\dot{\Sigma}_{\rm g}$ (denoted by open symbols), as 
observed in lower surface density galaxies  \citep{Chen+10}. In panel 2(b) we show the variation of the ratio of
 $\eta = \dot{\Sigma}_{\rm g}/\dot{\Sigma}_{\star}$, 
whereas the inset figure shows the variation of the star-formation rate alone;   for this purpose we have adopted
the \cite{Ken98} fit to the  star-formation rate density as a function of $(\Sigma_{\rm g}, \kappa^{-1})$
\EQ
\dot{\Sigma}_{\star} \approx 0.017\Sigma_{\rm g}\Omega = 0.012\Sigma_{\rm g}\kappa, 
\label{kenlaw}
\EN 
where $\Omega$ is the angular velocity and in the last equality, we have assumed $\kappa = \sqrt{2}\Omega$, as 
appropriate for a flat rotation curve. The symbols (open and filled) follow the convention of those in panel 2(a). 

We recall that large-scale outflows are most likely to occur in systems in which the star-formation 
rate, $\dot{\Sigma}_\star \geqslant 0.1\,{\rm M}_\odot\,{\rm yr}^{-1}\,{\rm kpc}^{-2}$. This implies that for each pair of values 
of $\Sigma_{\rm g}$ and $\kappa^{-1}$, massive outflows are expected for all runs except for: S50K30, S75K30, S50K20, 
S75K20, S100K20 and S150K20. Comparing with the gas mass loss rate plotted in panel 2(a), we find that runs S50K30, 
S75K30, S50K20 and S75K20 show very weak outflows (i.e. below the threshold value of 
$0.02\,{\rm M}_\odot\,{\rm yr}^{-1}\,{\rm kpc}^{-2}$).
Among the other two runs, the mass loss rates in S100K20 and S150K20 are on and above the threshold 
value respectively. Run S100K10  lies above the `Heckman limit,' but shows only a weak outflow. The 
ratio ($\eta$) of the gas mass loss rate and the star formation rate for runs with $\kappa^{-1} = 6.5$ and $10\,{\rm Myr}$ 
stays nearly constant, {\bf $\eta \approx 0.1$}. However, for $\kappa^{-1} = 20\,{\rm Myr}$, $\eta$ varies weakly with 
$\Sigma_{\rm g}$ for runs both above and below the Heckman limit. The trend for $\kappa^{-1} = 30\,{\rm Myr}$ is 
unclear due to our choice of simulation parameters of $\Sigma_{\rm g}$. 

\begin{figure*}[t!]
\centering
\includegraphics[width=2.0\columnwidth]{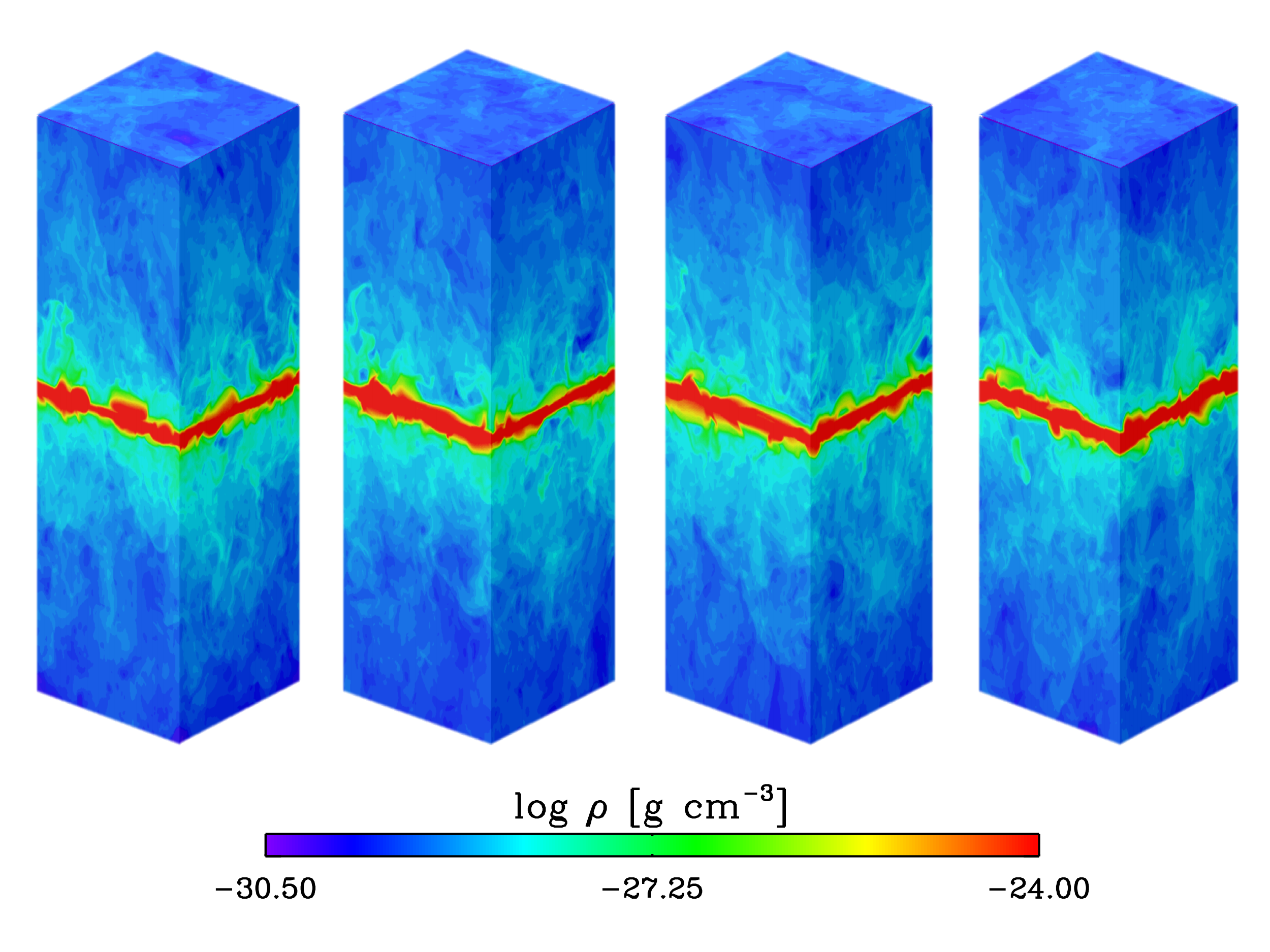}
\caption{(color online). A 3D rendering of the density in the run S250K10, showing the gradual evolution of
the outflow at four different times starting from $t = 312$ to $319\,{\rm Myr}$ (left to right). For clarity of 
the color contrast, we have restricted the density range from ${\rm log}\,\rho = -30.50$ to $-24\,[{\rm g\,cm^{-3}}]$. 
Plumes of gas move away from the disk mid plane and leave the simulation domain through the vertical boundaries.}
\label{3drend}
\end{figure*}

In Figure~\ref{3drend}, we show a three-dimensional rendering of the density at four different times, in a run 
with $\Sigma_{\rm g} = 250\,{\rm M_\odot\,pc^{-2}}$ and $\kappa^{-1} = 10\,{\rm Myr}$. These volume 
renderings show the evolution of an outflow with plumes moving gradually both upwards and downwards from 
the midplane, eventually leaving the simulation domain.
Note that the question of whether the ejected material is able to leave the simulation domain depends on 
whether the outflow velocity exceeds the escape velocity, which varies from simulation to simulation in our study.
Specifically, given the vertical extent of the box above the 
midplane and the gas surface density, the critical velocity required to escape the gravitational potential of the 
host galaxy is given by 
\EQ
v_{\rm es} = \sqrt{2\,g\,z_{\rm max}} \approx \left(\frac{103}{{\rm km\,s^{-1}}}\right)\,\left(\frac{\Sigma_{\rm g}}
{100\,{\rm M}_{\odot}\,{\rm pc^{-2}}}\right)^{1/2}\,\left(\frac{z}{\rm kpc}\right)^{1/2}, 
\label{ves}
\EN
where $g = 4\pi\,G\,\Sigma_{\rm g}$ is the acceleration due to gravity (assuming $\Sigma = 2\Sigma_{\rm g}$) and 
$z_{\rm max}$ is the vertical extent of the simulation volume. However,  since the vertical extent is three times the 
driving scale $R,$ which in turn would be related to the typical radial scale length, $R_{l}$ 
 (as $G\pi\Sigma\,R_{l}^{2} \approx \Omega^{2}\,R_{l}$, which implies $R_{l} \approx \pi\,G\Sigma/\Omega^{2} \approx R$)
in a more complete model disk evolution, 
we can reasonably expect that in a real disk as material travels to heights approaching $z_{\rm max}$, $g(z)$ would 
be falling rapidly, resulting in escape velocities similar to those in our simulations.

\begin{figure}[ht!]
\centering
\includegraphics[width=1.0\columnwidth]{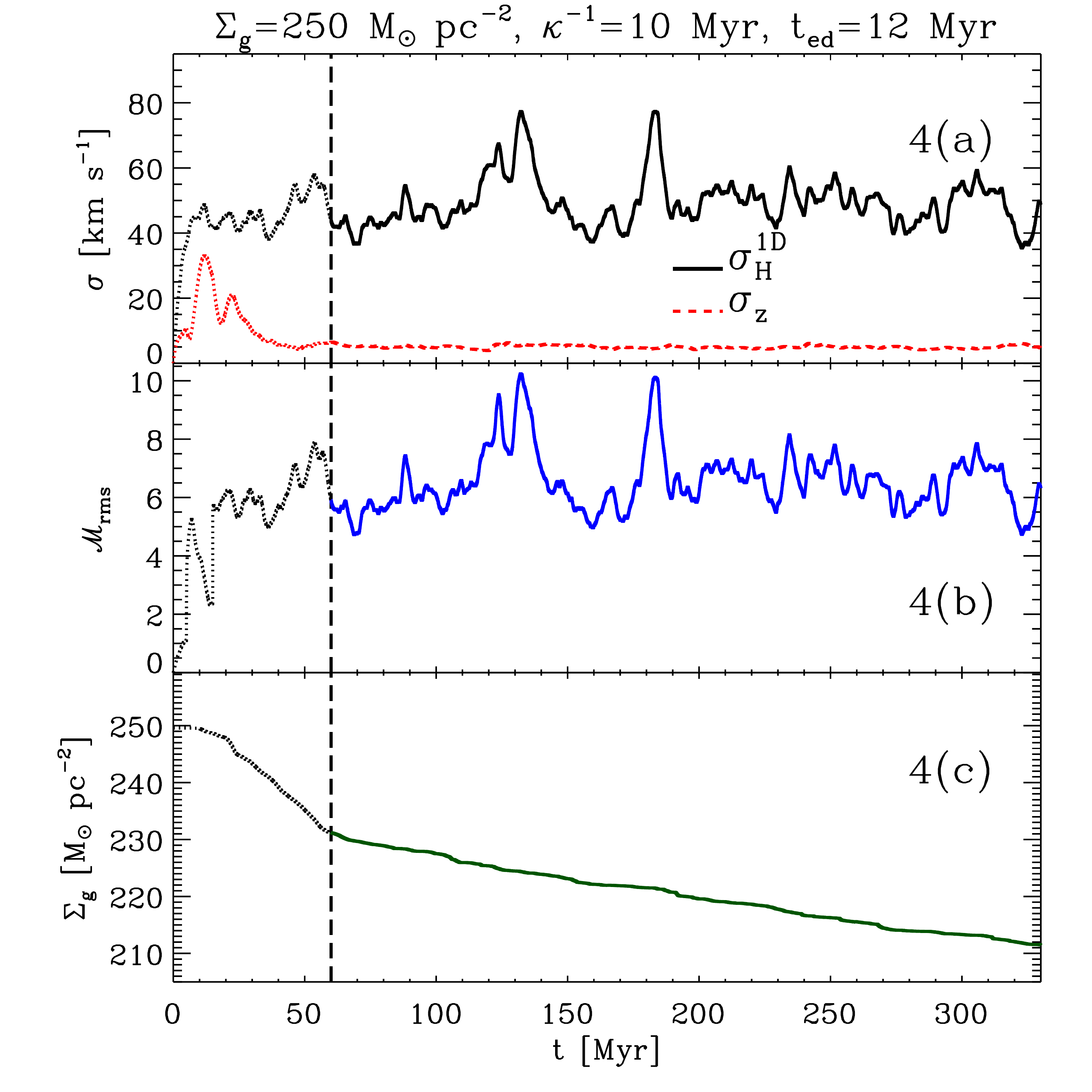}
\caption{(color online). 
Time evolution of the mass-weighted one-dimensional
horizontal velocity dispersion (black) and the vertical velocity dispersion (red) in panel 4a), the 
rms Mach number (blue) in panel 4b), and the gas mass surface density (dark green) in panel 4c) 
for run SK250K10 with $\Sigma_{\rm g} = 250\,{\rm M_\odot\,pc^{-2}}$, $\kappa^{-1} = 10\,{\rm Myr}$, 
and $t_{\rm ed} = 12\,{\rm Myr}$. The dashed vertical line depicts the duration of the initial transient 
phase. The values of $\sigma^{\rm 1D}_{\rm H}$, $\sigma_{\rm z},$ and $\dot{\Sigma}_{\rm g}$ 
presented in Table~\ref{sumsim} are thus computed from 60 Myr onwards. Notice that there is 
significant mass loss at times well after the initial transient phase.}
\label{ts500k10}
\end{figure}

\begin{figure}[ht!]
\centering
\includegraphics[width=1.0\columnwidth]{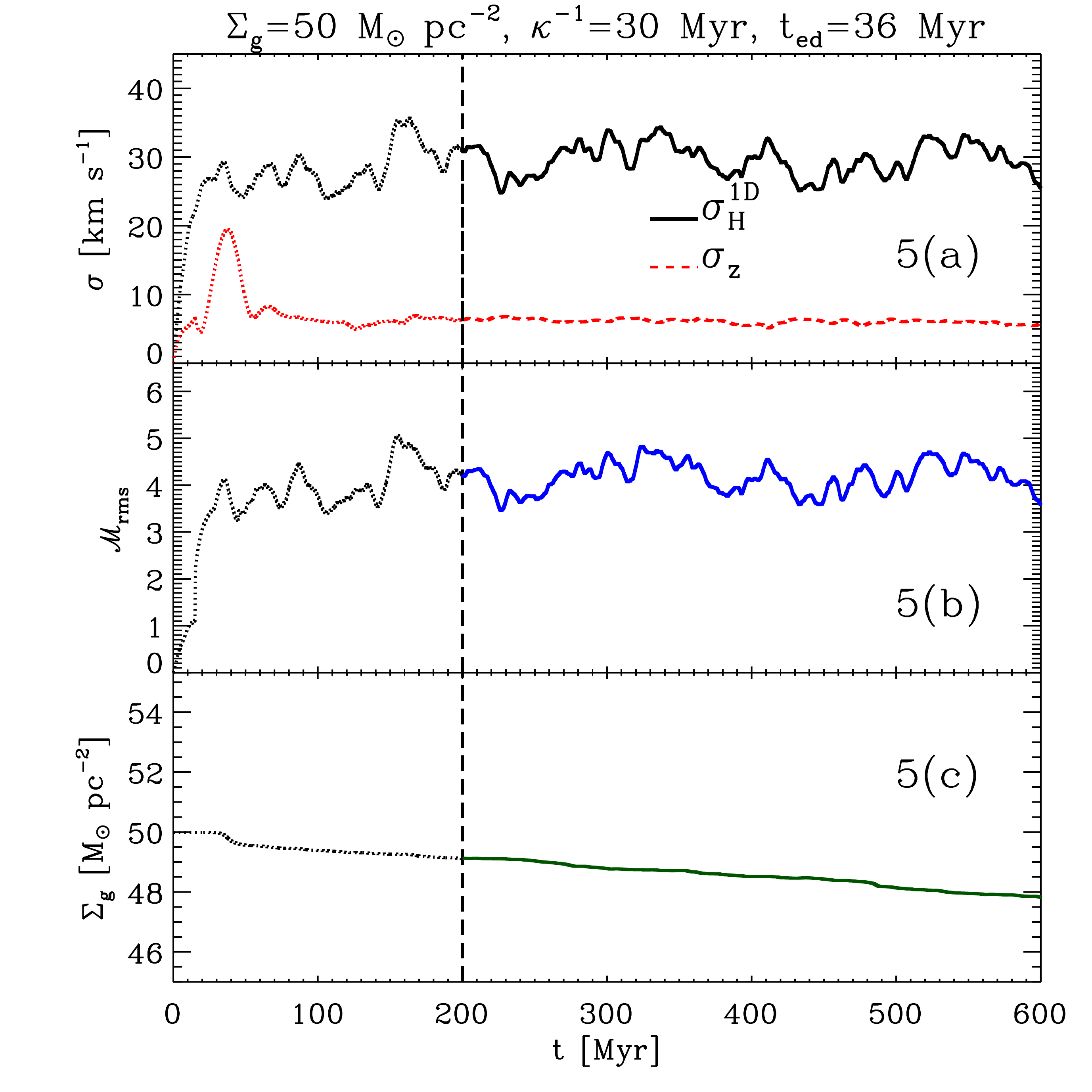}
\caption{(color online).
Same as in Fig.~\ref{ts500k10} for run SK50K30 with
$\Sigma_{\rm g} = 50\,{\rm M_\odot\,pc^{-2}}$, $\kappa^{-1} = 30\,{\rm Myr},$ and 
$t_{\rm ed} = 36\,{\rm Myr}$. In this case these is negligible mass loss over a 400 Myr long period.}
\label{ts100k30}
\end{figure}

\subsection{Time Series, Vertical structure, and Thermal Phases}
\label{tseries_2dplots}

Before we delve further into the dependence of the outflows on the disk properties, we show 
the time evolution of two runs that illustrate the range of behaviors seen in our simulations.
Figure~\ref{ts500k10}  shows the evolution of
S250K10, a high surface density, rapidly-rotating case with $\Sigma_{\rm g} = 250 {\rm M_\odot\,pc^{-2}}$ 
and $\kappa^{-1} = 10$ Myr, and Figure~\ref{ts100k30} show the evolution of 
S50K30, a moderate surface density, slowly-rotating case with 
$\Sigma_{\rm g} = 50 {\rm M_\odot\,pc^{-2}}$ and 
$\kappa^{-1} = 30$ Myr. In both figures, we plot the mass-weighted one-dimensional horizontal velocity dispersion, 
{\bf $\sigma^{\rm 1D}_{\rm H} = \sigma_{\rm H}/\sqrt{2}$}, the vertical velocity dispersion, $\sigma_{\rm z}$,
 the mass-weighted rms Mach number, and the gas mass surface density.  
In both of these runs, the initial time evolution is marked by a transient phase, which lasts for $60$ Myr and 
$200$ Myr respectively (denoted by the dashed vertical lines), after which the system reaches a steady state.
In this stage, the action of gravity in the absence of turbulent support causes 
the material to move downwards towards the mid plane and become compressed. Then as turbulence 
develops, pressure support also increases, puffing the layer back up, and causing a rapid expansion 
that drives a small fraction of the material out of the simulation volume. During these rearrangements, our results 
are strongly dependent on the particulars of our initial conditions, and so we avoid making measurements or 
drawing conclusions from this phase of the simulations. 

On the other hand, after the initial phase has passed, the simulations settle into a quasi-steady state that depends 
almost purely on $\Sigma_{\rm g} $ and $\kappa.$  We thus evaluate the time-averaged values in Table~\ref{sumsim} 
only in this quasi-steady state. By this time, the medium in the S250K10 run settles into a roughly constant 
mass-weighted 1D horizontal velocity dispersion of $\bar{\sigma}_{\rm H}^{\rm 1D} = 50\,{\rm km\,s^{-1}}$ while the 
medium in the SK50K30 run reaches $\bar{\sigma}_{\rm H}^{\rm 1D}  = 30\,{\rm km\,s^{-1}}$. While the higher surface 
density run is somewhat hotter than the moderate surface density case, in both runs the sound speed is well below 
the average velocity dispersion.  This implies that the turbulent motions are supersonic in nature. 
In fact, the mass weighted rms Mach number $\approx 6$ for the higher surface density run and 
$\approx 4$ in the lower surface density run, signifying the occurrence of strong shocks in both runs.
Thus, although the driving in these simulations is purely 
horizontal, it also causes substantial pressure and density perturbations, which vary both horizontally and vertically. 
These lead in turn to vertical velocity fluctuations, which are roughly constant in time at 
$\bar{\sigma}_{\rm z} = 5.5\,{\rm km\,s^{-1}}$ in the S250K10 case and $\bar{\sigma}_{\rm z}  = 6\,{\rm km\,s^{-1}}$ in 
the S50K30 case. This shows that the presence of supernova energy input is not required for high-density disks to 
develop significant turbulent motions perpendicular to the plane of the galaxy, provided large-scale self-gravitating 
instabilities are able to maintain horizontal motions of several tens of ${\rm km\,s^{-1}}$.
Quite remarkably, as Table~\ref{sumsim} shows, the value of the time averaged mass-weighted vertical velocity 
dispersion is consistently within the range from $5 - 6.3\,{\rm km\,s^{-1}}$ in our simulations, which we discuss in 
more detail below.

\begin{figure*}[ht!]
\centering
\includegraphics[width=1.8\columnwidth]{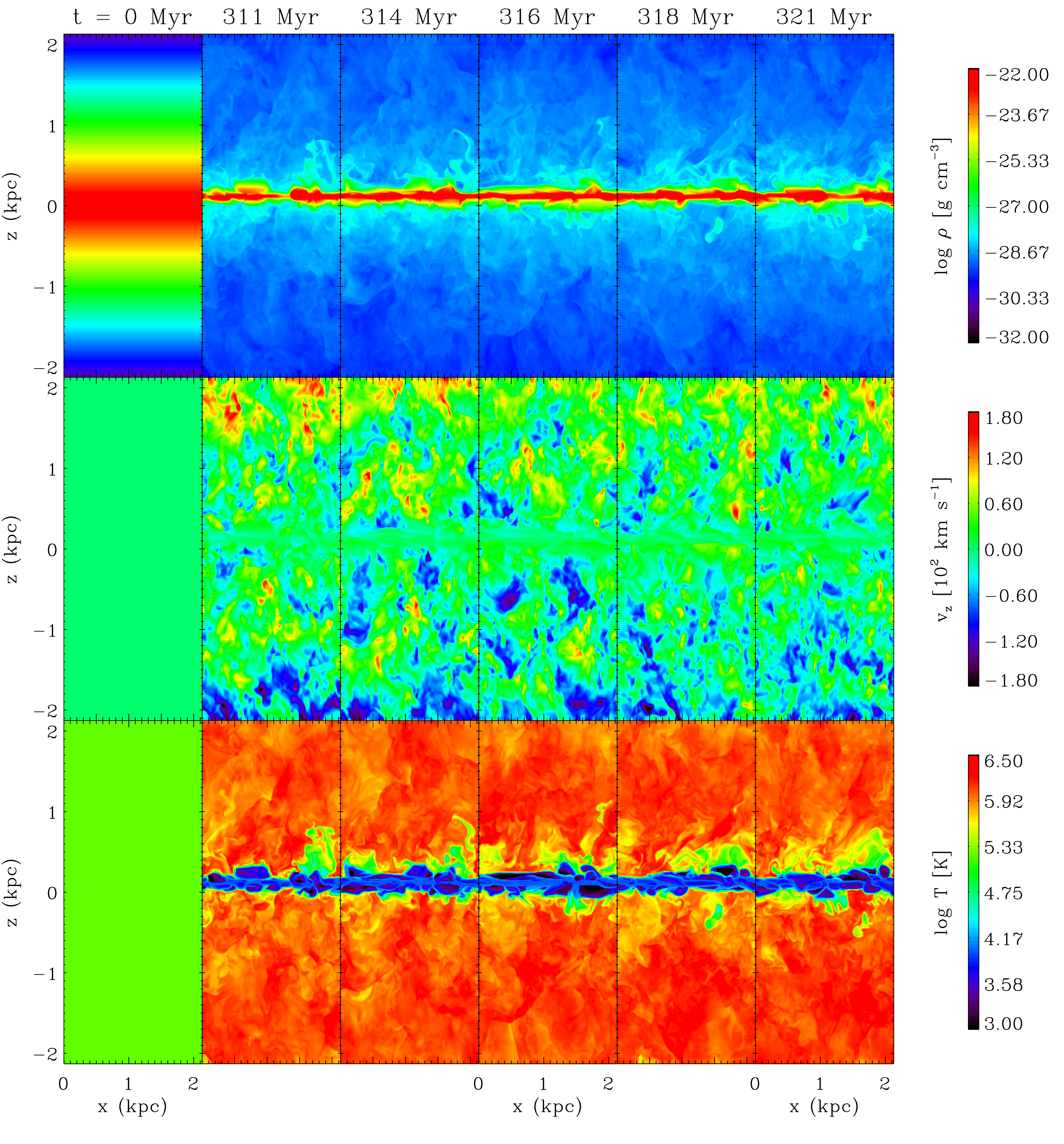}
\caption{(color online). Two-dimensional slices in the $x-z$ plane showing the 
time evolution of the logarithm of the density (upper row) and the vertical component 
of the velocity (middle row) and the logarithm of the temperature (lower row) in a 
local patch of the turbulent ISM for a run with $\Sigma_{\rm g} = 250\,{\rm M_\odot\,pc^{-2}}$, 
$\kappa^{-1} = 10\,{\rm Myr}$ and $t_{\rm ed} = 12\,{\rm Myr}$. The total vertical extent 
of the simulation domain is $2z = 4.26\,{\rm kpc}$ while the horizontal extent is 
$x=y=2.13\,{\rm kpc}.$ }
\label{s500k10}
\end{figure*}

A closer look at these plots also reveals that for the S250K10 run, a distinct outflow occurs throughout the full 
quasi-steady state evolution (see panel 4c), persisting over many eddy-turnover times until the simulation is 
terminated at $t =  330\, {\rm Myr.}$  This is also evident from the two-dimensional $x-z$ slices of the density, 
vertical component of the velocity, and temperature in Figure~\ref{s500k10}. Here we see that, in course 
of the evolution, the combined action of turbulent heating, radiative cooling, and gravitational collapse leads to 
strong density and temperature contrasts throughout the simulation. Furthermore, the slices of the density plotted 
in the first row of this figure, together with the vertical velocity plotted in the second row, clearly show an outflow 
of material.

\begin{figure*}[ht!]
\centering
\includegraphics[width=1.8\columnwidth]{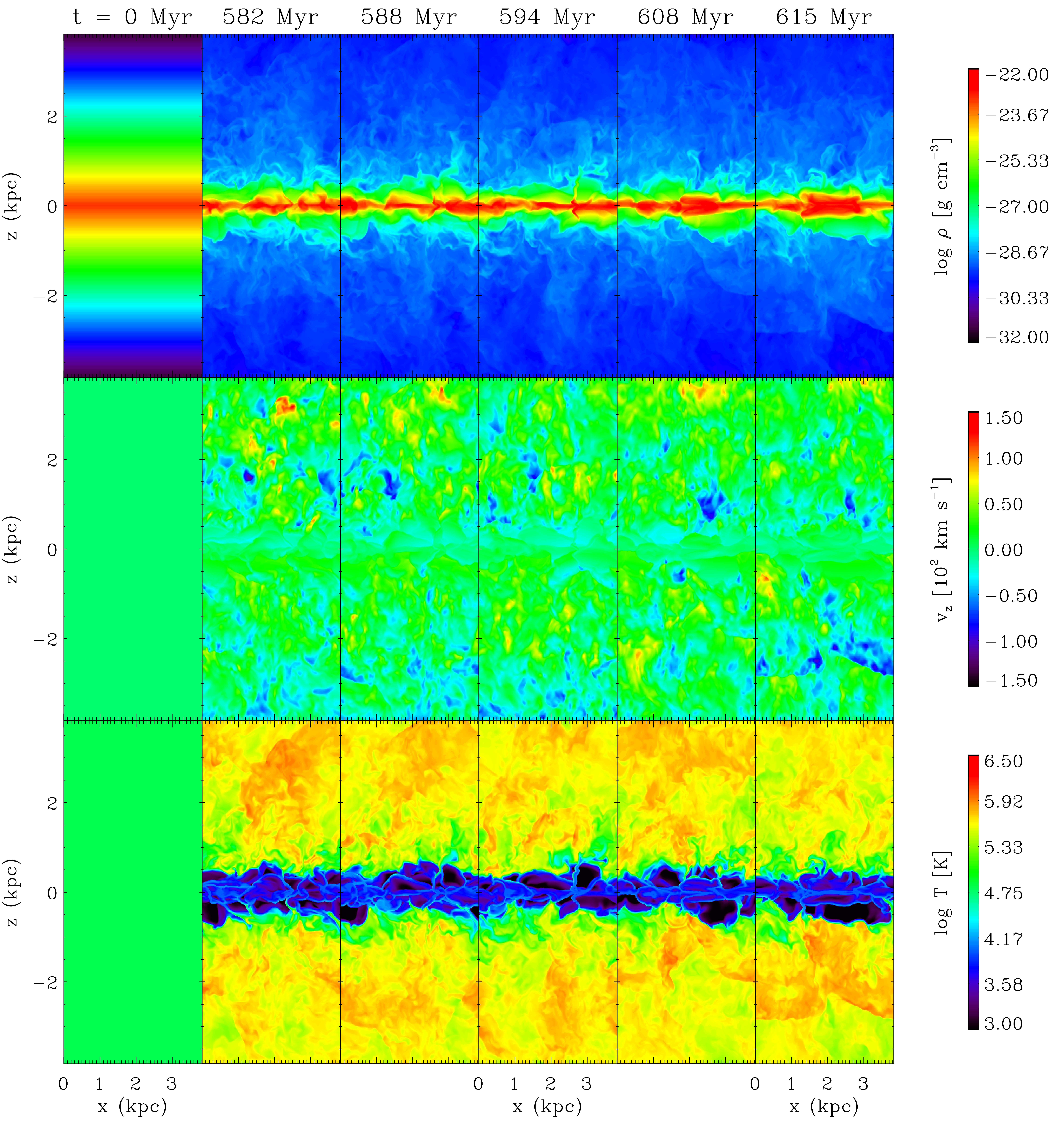}
\caption{(color online). Two-dimensional slices in the $x-z$ plane showing the 
time evolution of the logarithm of the density (upper row) and the vertical component 
of the velocity (middle row) and the logarithm of the temperature (lower row) in a local 
patch of the turbulent ISM for a run with $\Sigma_{\rm g} = 50\,{\rm M_\odot\,pc^{-2}}$,  
$\kappa^{-1} = 30\,{\rm Myr}$ and $t_{\rm ed} = 36\,{\rm Myr}$. The total vertical 
extent of the simulation domain is $2z = 7.66\,{\rm kpc}$ while the horizontal 
extent is $x=y=3.83\,{\rm kpc}.$}
\label{s100k30}
\end{figure*}

At this point, it is worth describing in greater detail how such outflows are generated. To answer this question, we 
return to the  total atomic, ionic, and Bremsstrahlung cooling curve in Figure~\ref{cool_curve}. The temperature 
slices in Figure~\ref{s500k10} show that by $t=311\,{\rm Myr}$, the initial constant temperature distribution has 
evolved into a multiphase structure. In particular, material in certain regions close to the mid plane is heated to 
$T\approx 10^{6}\,{\rm K}$ and higher, while some patches of gas are still close to a few times $10^{5}\,{\rm K}$.  
At the same time, Figure~\ref{cool_curve}, shows that beyond $T\approx 2.5\times10^{5}\,{\rm K}$, the radiative 
cooling rate decreases. This implies that these regions would take many dynamical times to cool to the mean 
temperature of the medium. But by then, successive heating resulting from turbulent driving would further heat the 
material. 

This behavior is also expected from a linear stability analysis of a temperature perturbation in a medium in which 
the cooling rate per unit mass is $\propto \rho \Lambda(T)$, and the heating rate per unit mass is roughly constant, 
as is the case with turbulent heating \citep{PP09}.
In this case, the the condition for such a perturbation to grow exponentially (at constant pressure) is simply
\EQ
\frac{\partial \ln \Lambda}{\partial \ln T} < 1,
\EN
\citep{Field65,Def70,MSQP12}. In a steadily-heated thermally-unstable medium, high density regions condense 
and cool, while low density regions expand and heat. When $\sigma$ is large,  
the temperature is high and  ${\partial \ln \Lambda}/{\partial \ln T} \lesssim 0$ 
throughout much of the medium.  This will result in an {\it unstable} state in which rapid, runaway heating 
of the hot, low density medium will lead to the significant removal of the gas from the simulation domain 
\citep[see also][]{SMQP12, SGP12}. Thus, from  $t=311\,{\rm Myr}$ onwards, Figure~\ref{s500k10} shows 
the continuing impact of this hot gas, which pushes its way outward both through the top and bottom boundaries.

\begin{figure*}[ht!]
\centering
\includegraphics[width=2.0\columnwidth]{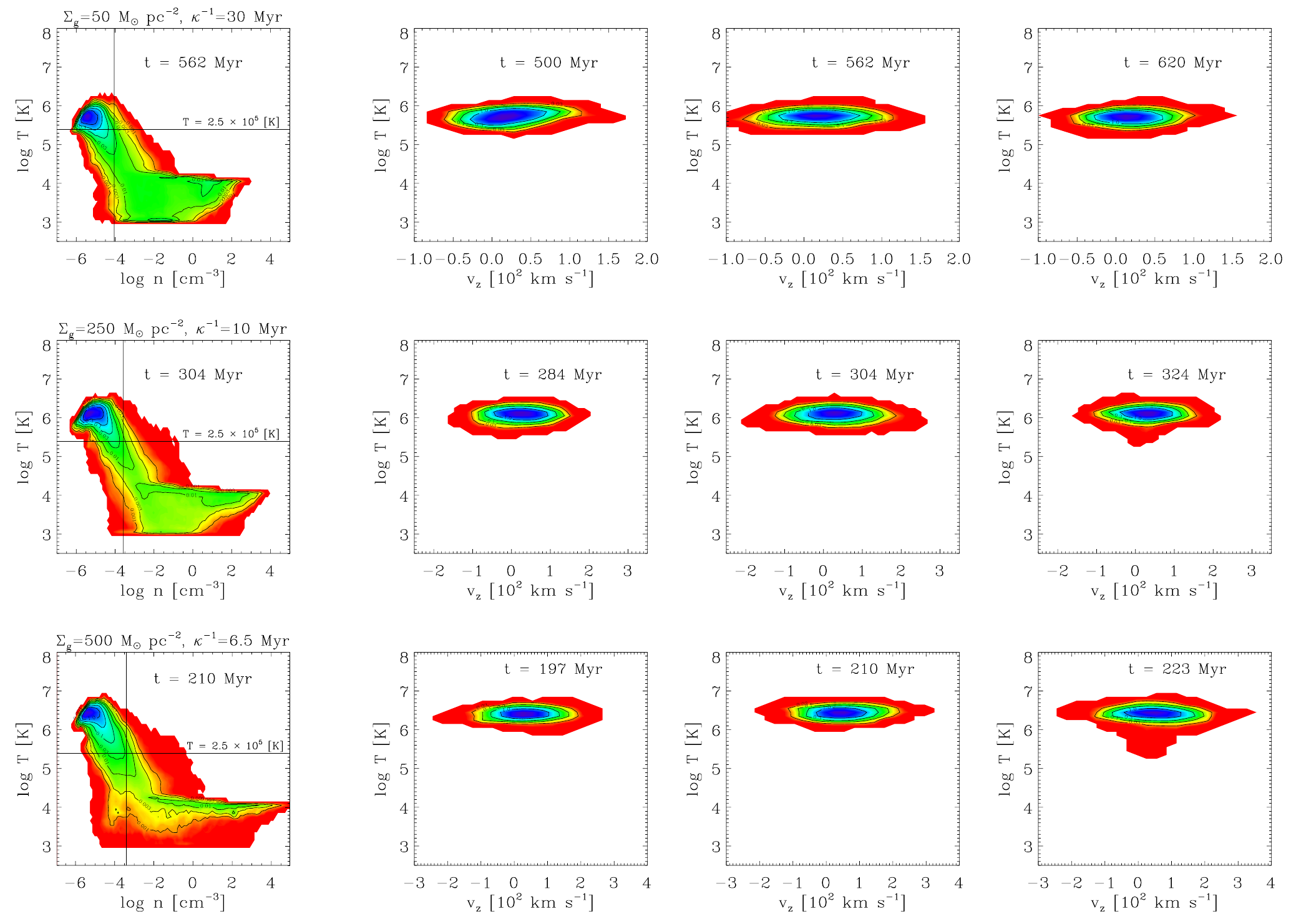} 
\caption{(color online). 
Volume-weighted phase diagrams of the temperature and the number 
density of all the gas (first column) and the mass-weighted temperature and the vertical component of the velocity PDFs 
of the high-latitude gas at different times (from second to the fourth column) for runs: S50K30 with 
$\sigma_{\rm H}^{\rm 1D}=30$ km/s and  $v_{\rm esc}=143$ km/s (top row), S250K10 with 
$\sigma_{\rm H}^{\rm 1D}=50$ km/s and $v_{\rm esc}=238$ km/s  (middle row) and S500K6.5 with 
$\sigma_{\rm H}^{\rm 1D}=68$ km/s $v_{\rm esc}=309$ km/s  (bottom row). The $T-n$ phase diagrams
show $\log_{10}$ contours of the probability density function, 
computed over the entire simulation domain, and normalized to 1 when integrated over $\ln T$ and $\ln n$. 
The  $T-v_{\rm z}$ phase plots show $\log_{10}$ contours of the probability density function, computed over the range 
$(z_{\rm max}/2 - z_{\rm max})$ for each of the three runs, and normalized to 1 when integrated in $\ln T$ 
and $v_{\rm z}$ in km/s. In all panels two contour lines are plotted per decade.
}
\label{phaseplots}
\end{figure*}

In contrast, for a slowly-rotating, moderate surface density disk, the temperature slices in Figure~\ref{s100k30} show 
that the majority of the material away from the midplane is at temperatures $\leqslant 2.5\times 10^{5}\,{\rm K}$. 
In this phase, ${\partial \ln \Lambda}/{\partial \ln T} \geq 1$ and the cooling time is roughly constant or dropping as 
a function of temperature, leading to a stable configuration with less significant mass loss through the 
vertical boundaries.   

Such multiphase outflowing material is also detected in observed high surface density 
starbursting galaxies, which contain both $10^{7}-10^{8}\,{\rm K}$ emitting material \citep{Martin99, SH07, SH09} 
and $\approx 10^{4}\,{\rm K}$ material detectable in a variety of optical and near UV absorption and emission 
lines \citep{Pettini+01, Tremonti+07}. Certainly, in real galaxies, much of the hottest of these phases is in fact material 
from supernovae. On the other hand, the presence of $\approx10^{6}\,{\rm K}$ gas may provide pockets in which the 
cooling of material shocked by supernova remnants is minimal, making such galaxies particularly prone to outflows. 
Or in other words, gravitationally-driven turbulence is likely to both drive a baseline outflow as well as provide an 
ISM distribution in which supernova driving is particularly efficient. 
 
Figure~\ref{phaseplots} shows the phase distributions of the gas for three different runs, including  
the temperature-density distribution over the whole simulation volume and the temperature-velocity distribution
of the gas at high latitudes. In the $T-n$ plots, $T = 2.5 \times 10^{5}\,{\rm K}$ is denoted by the horizontal line, 
above which the cooling rate starts to decrease. The vertical lines in the plots denote the densities at which
the cooling time at $T = 2.5\times 10^{5}\,{\rm K}$ is equal to the eddy turnover time. To the left of this line 
and for $T > 2.5\times 10^{5}\,{\rm K}$, the cooling time is greater than $t_{\rm ed},$ while to the right, 
$t_{\rm cool} < t_{\rm ed}$. In all three plots, we find the existence of a horizontal, high density tail at 
$T \approx 10^{4}\,{\rm K}$; a feature commonly seen in phase diagrams in ISM simulations. 
In the S50K30 case, corresponding to a slowly rotating, moderate surface density galaxy, 
a majority of material is below the threshold temperature of $2.5\times 10^{5}\,{\rm K}$. This run has only negligible 
mass loss, and the $T-v_{\rm z}$ plots show that the vertical component of the velocity corresponding to the gas 
near the top boundary is $50-100\,{\rm km\,s^{-1}},$ as compared to the escape velocity of 143 km/s.

On the other hand, in both S250K10 with $\bar{\sigma}_{\rm H}^{\rm 1D}=50$ ${\rm km\,s^{-1}}$ and S500K6.5 
with $\bar{\sigma}_{\rm H}^{\rm 1D}=68$ ${\rm km\,s^{-1}}$, more of the low density gas above the disk mid plane is 
at considerably higher temperatures than the temperature at which the cooling time is equal to the eddy turnover time. 
Due to the longer cooling time, this gas becomes thermally unstable and leads to a rapid, runaway heating
resulting in an eventual expulsion of the gas from the host galaxy, dragging along additional material with it. 
Thus, such high density, faster spinning disks are more likely to host galaxy outflows than their slow rotating less 
denser counterparts. The mass-weighted phase plots of the temperature and the vertical component of the velocity 
for S250K10 and S500K6.5 show that the outflowing gas is made up of a combination of phases with most of the 
mass in the hot phase at $T\approx 10^{6}-10^{7}\,{\rm K}$ which also drags along some relatively low temperature 
gas as it escapes the galaxy. In both cases the vertical velocities range from $\approx 100\,{\rm km\,s^{-1}}$ to close 
to the escape velocity, which is 238 ${\rm km\,s^{-1}}$ in the S250K10 case and 309 ${\rm km\,s^{-1}}$ in the 
S500K6.5 case.

\subsection{Turbulent Parameters and Vertical Scale Heights} 

\begin{figure*}[t!]
\centering
\includegraphics[width=2.0\columnwidth]{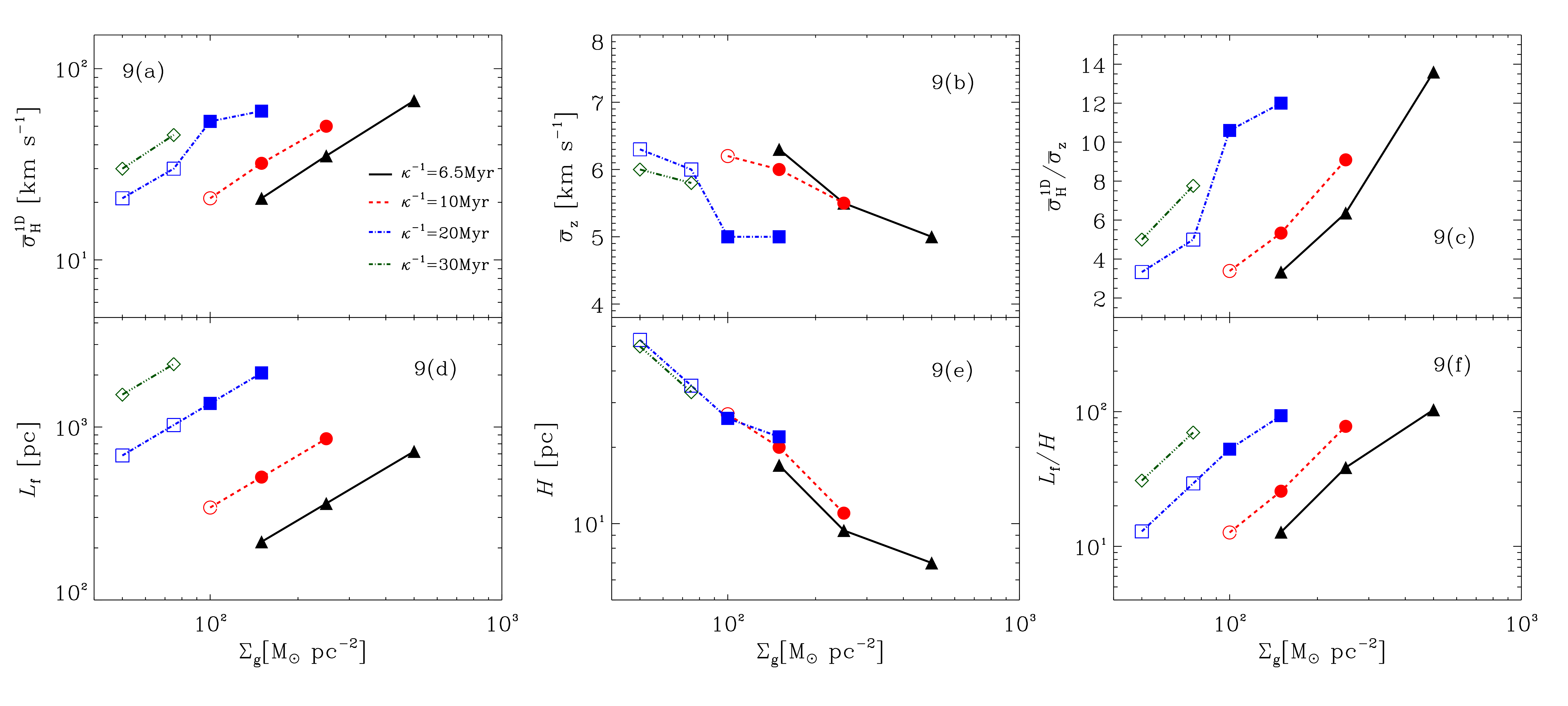} 
\caption{(color online).
The different panels shows the variation of the following quantities as a function of the gas surface density 
for different ranges of the epicyclic frequency. Panel 9(a) - computed one-dimensional horizontal velocity 
dispersion, panel 9(b) - vertical velocity dispersion, panel 9(c) - the ratio of the horizontal to the vertical 
velocities, panel 9(d) - the horizontal forcing scale, $L_{\rm f}$, panel 9(e) - the scale-height, $H$ 
as computed in equation~(\ref{vertscale}) and the ratio of $L_{\rm f}/H$ in panel 9(f). Note that while 
$\sigma_{\rm H} \propto \Sigma\,\kappa^{-1}$ and $L_{\rm f} \propto \Sigma\,\kappa^{-2}$ are both controlled 
by the imposed turbulent driving, the rest of the quantities represent the natural  response to horizontally 
driven motions.
}
\label{veldis}
\end{figure*}

To further explore the connection between the mass loss rate and the structure of the disk, we measured 
how  turbulent velocities and vertical scale-heights vary with our  input parameters, $\Sigma_{\rm g}$ 
and $\kappa.$  In Figure \ref{veldis}, the filled (open) symbols follow the classification of disks hosting
outflows above (below) $0.02\,{\rm M}_\odot\,{\rm yr}^{-1}\,{\rm kpc}^{-2}$, similar to those in panel 
2(a) of Figure~\ref{starform}. Comparing the mass loss rate with the one-dimensional horizontal velocity 
dispersion, we find that with the exception of S75K30, strong outflows are ubiquitous 
in systems where $\bar{\sigma}^{\rm 1D}_{\rm H} \geqslant 32\,{\rm km\,s^{-1}}$.  
This is in agreement with 
the earlier numerical \citep{SGP12} and analytical predictions \citep{Scann13}, which argued that 
provided turbulent velocities can be maintained
above this threshold value (about $35\,{\rm km\,s^{-1}}$ in 
their paper), the material is shocked into the thermally unstable regime, leading to runaway heating and 
eventual expulsion from the galaxy.
A compilation of a large sample of high-redshift data by \citet{Genzel+11} also shows that galaxies with 
$\dot{\Sigma}_{\star}\geqslant \dot{\Sigma}_{\star}^{\rm cr}$ have 
$\sigma_{\rm H\alpha}^{\rm 1D}\geqslant 35\,{\rm km\,s^{-1}}$. 
On the other hand, the vertical velocity dispersion varies 
only weakly across the entire parameter space with a mean value of $5.7\,{\rm km\,s^{-1}}$.

We note here that the vertical velocity dispersion arises only in response to the horizontal turbulent motions. 
Even though only the horizontal turbulent velocity is stirred externally at a velocity proportional to $\Sigma_{\rm g}$
to model gravitational driving in an unstable disk, panel 9(b) shows that the vertical velocity dispersion decreases 
only weakly with increasing gas surface density with $\sigma_{\rm z} = 5.0 - 6.3\,{\rm km\,s^{-1}}$. Intriguingly, 
the small and nearly constant value of $\sigma_{\rm z}$ with the gas surface density is quite similar to the results 
from simulations in which turbulence is driven by supernovae, with a star formation rate set self-consistently by 
local gravitational collapse \citep{KKO11, SO12, KOK13}. 
However, for turbulence driven by gravitational instability, the plot of $\bar{\sigma}^{\rm 1D}_{\rm H}/\bar{\sigma}_{\rm z}$ 
in panel 9(c) suggests that turbulent velocities become increasingly {\it anisotropic} with increasing gas surface density. 
For a constant $\Sigma_{\rm g}$, the anisotropy increases for slowly rotating disks. 

The variation of the imposed horizontal forcing scale, $L_{\rm f}\approx R \propto \Sigma_{\rm g}/\kappa^2$ 
is shown for reference in panel 9(d). At constant $\kappa^{-1}$, the forcing scale is directly proportional to the gas 
surface density, while at constant $\Sigma_{\rm g}$, slower rotating disks have a larger forcing scale. The vertical 
scale-height, $H$ shown in the panel 9(e) is measured directly from our simulations as the mass-weighted mean 
height of the gas, 
\EQ
H \equiv \int_{z_{\rm min}}^{z_{\rm max}} |z|\rho\,dV / \int_{z_{\rm min}}^{z_{\rm max}}\rho\,dV. 
\label{vertscale}
\EN
Since $\sigma_{\rm z}$ remains more or less constant with $\Sigma_{\rm g}$, and $H \propto \sigma_z^2/(G \Sigma)$, 
this results in a decrease of the scale height with increasing $\Sigma_{\rm g}$ independent 
of the value of $\kappa^{-1}$. On the other hand, for a given value of $\Sigma_{\rm g}$, the scale height is 
identical for different values of $\kappa^{-1}$.

Finally, we find that the ratio of the forcing scale to the disk scale-height in panel 9(f) again increases with 
increasing $\Sigma_{\rm g}$ and for a given surface density of the disk.  It is also larger for more slowly 
rotating disks.  For all models, we obtain $L_{\rm f}/H > 10$. Thus, even in cases with large $H$, the forcing is still 
at large scales compared to the disk thickness as would be expected for gravitationally-driven instabilities.

\subsection{Timescales} 

The vertically stratified medium in our simulations is constantly heated by turbulent driving 
and simultaneously cooled by atomic, ionic, and Bremsstrahlung, optically-thin radiative cooling. In the context of 
galaxy clusters, earlier works by \citet{MSQP12} and \citet{SMQP12} showed that such a medium only develops 
a multiphase structure if the free-fall timescale is longer than the average cooling time, where 
the local cooling time is given by equation~(\ref{tcool}). Note, however, that \citet{MSQP12} employed  
a cooling function that was dominated by  thermal bremsstrahlung  while \citet{SMQP12} included a cooling function 
similar to ours \citep{SD93}, but focused on a higher range of temperatures. More importantly, cluster turbulence is 
incompressible \citep[e.g.,][]{Churazov+12, SF12}, in contrast to our simulations, in which turbulence is supersonic, 
and the majority of the energy is kinetic rather than thermal. 

\begin{figure*}[t!]
\centering
\includegraphics[width=2.0\columnwidth]{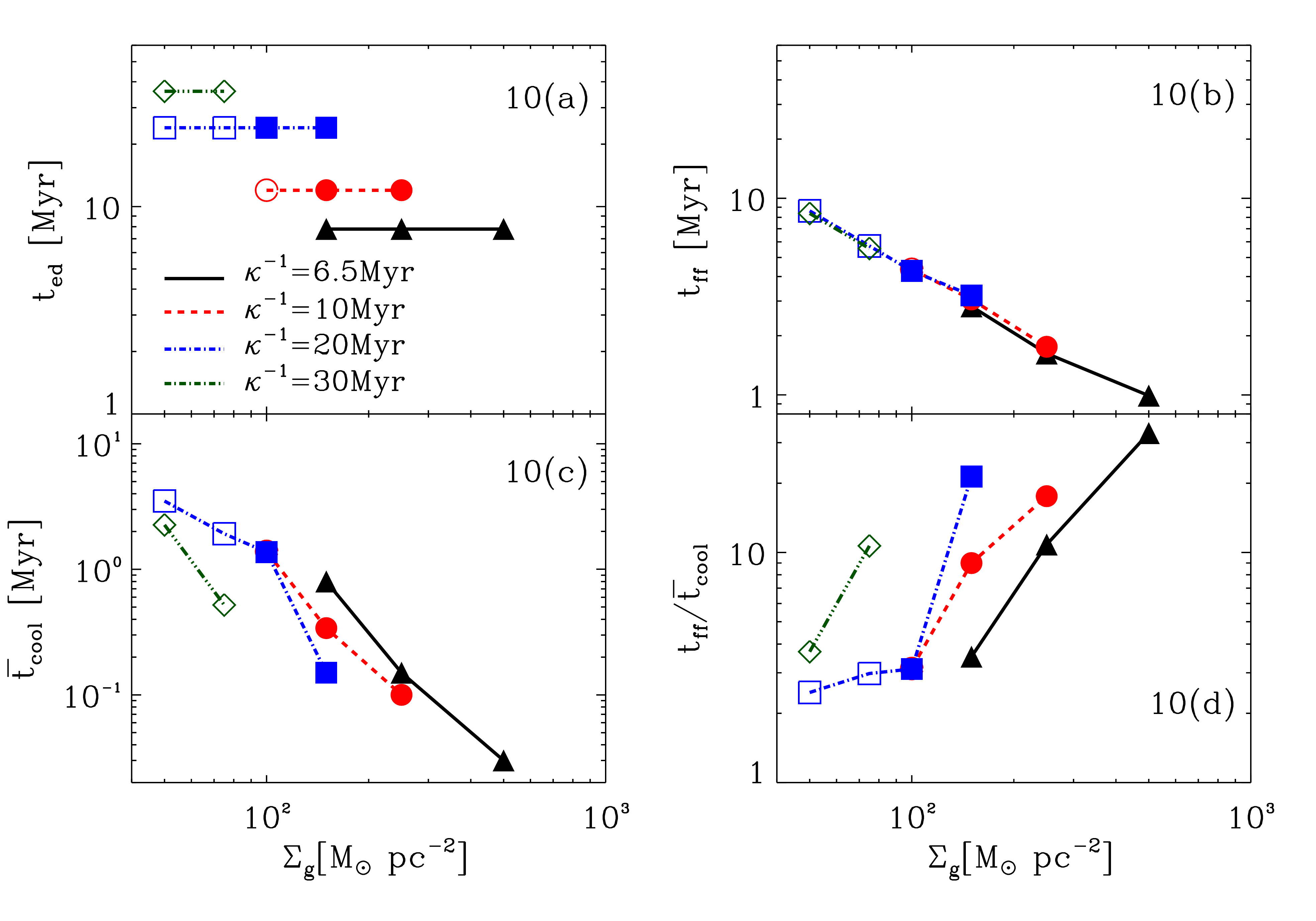} 
\caption{(color online). Variation of the eddy-turnover time (panel 10a), the free-fall time (panel 10b), the average 
cooling time (panel 10c), and the ratio of the free-fall to the cooling time (panel 10d) with the gas surface density 
for simulations with different values of $\kappa^{-1}$. 
}
\label{tscales}
\end{figure*}

In Figure~\ref{tscales}, we show the variation of the free-fall time, the eddy-turnover time, the cooling time, and 
the ratio of the free-fall to the cooling timescale in our simulations as a function of surface density and rotation rate. 
Including constant factors, the eddy-turnover time in our simulations is, $t_{\rm ed} \approx 1.2\,\kappa^{-1},$ 
i.e., independent of $\Sigma$.
However, we must analyse our simulation results to obtain the free-fall timescale defined as
\EQA
t_{\rm ff} &\equiv&
\left(\frac{H}{2\pi\,G\,\Sigma_{\rm g}}\right)^{1/2} \\ \nonumber
&=& 5.9\,{\rm Myr}\,\left(\frac{H}{100\,{\rm pc}} \right)^{1/2} 
\left( \frac{\Sigma_{\rm g}}{100\,{\rm M}_\odot {\rm pc}^{-2}}\right)^{-1/2},
\label{tff}
\ENA
where 
we quote $t_{\rm ff}$ in the plane in which $z = H,$ the vertical scale height defined in eq.\  (\ref{vertscale}). 
Using the values of $H$ from 
Figure~\ref{veldis} in the above equation, we find that $t_{\rm ff}$ steadily decreases with $\Sigma_{\rm g}$ 
for a given value of the epicyclic frequency. Moreover, at a constant $\Sigma_{\rm g}$, the free-fall time scales 
are similar due to identical values of the scale height.
The ratio of the free-fall time to the eddy-turnover time can also be written as
\EQ
\frac{t_{\rm ff}}{t_{\rm ed}} = \frac{\kappa}{1.2\,\sqrt{2\pi\,G}}\left(\frac{H}{\Sigma_{\rm g}}\right)^{1/2} 
= 0.83\left(\frac{H}{R}\right)^{1/2},
\label{tffoverted}
\EN
where in the last equality we have substituted $\Sigma_{\rm g} = \kappa^{2}\,R/2\pi\,G,$ where $R$ is the size of the 
eddy at the forcing scale $L_{\rm f}$. Since the scale-height $H \ll L_{\rm f}$ in our simulations (see figure~9f), the 
free-fall timescale is in general smaller than the eddy-turnover time. This is also evident by comparing panels 10 (a) 
and (b) in the above figure. 
Equation~(\ref{tffoverted}) can also be expressed as the ratio of the vertical to horizontal 
velocity dispersion by noting that $\sigma_{\rm z} \approx 
(2\pi\,G\,\Sigma_{\rm g}\,H)^{1/2}$ and 
$R = \sigma_{\rm H}/\kappa$ such that
\EQ
\frac{t_{\rm ff}}{t_{\rm ed}} = \frac{0.8\,\sigma_{\rm z}}{(2\pi\,G\,\Sigma_{\rm g})^{1/2}}\left(\frac{\kappa}{\sigma_{\rm H}}\right)^{1/2}
= \frac{0.8}{Q^{1/2}}\left(\frac{\sigma_{\rm z}}{\sigma_{\rm H}}\right), 
\label{teff_vel}
\EN
where $Q = 2\pi\,G\,\Sigma_{\rm g}/\kappa\,\sigma_{\rm H}$ is the Toomre parameter. Thus for $Q=1$, the ratio of the 
free-fall to the eddy-turnover time can be interpreted as the ratio of the vertical to the horizontal velocity dispersions 
and therefore as a measure of the isotropy of the velocities. 

Using the scale height and the vertical velocity dispersion, one could construct a vertical mixing time scale, 
$t_{\rm mix} = H/\sigma_{\rm z} = \sigma_{\rm z}/2\,\pi\,G\,\Sigma_{\rm g}$, where we have substituted, 
$H \approx \sigma_{\rm z}^{2}/2\,\pi\,G\,\Sigma_{\rm g}$. Using the values of $\sigma_{\rm z}$ and $\Sigma_{\rm g}$
from Table~\ref{sumsim}, we find that the mixing time scale is similar to the free fall time scale with high surface 
density, rapidly rotating disks having smaller mixing time scales compared to low surface density, slowly rotating 
disks.

In Panel 10(c) we show the mass-weighted average cooling time, $\bar{t}_{\rm cool}$, defined as the ratio of 
the total thermal energy to the total cooling rate. 
We find that the cooling times are particularly short in high surface density disks. We note here that in our simulations, 
the turbulence (modeling motions driven by gravitational instabilities) is converted into thermal energy and then radiated 
away. Thus one could in principle also estimate the cooling time as the time taken to radiate away both the thermal 
energy and the kinetic energy from the system. While both these times are similar in a subsonic medium such as studied 
by \citet{MSQP12} and \citet{SMQP12}, in a supersonic medium the time to radiate away the total energy would be 
much longer than the time to radiate only the thermal energy.

The ratio of the free-fall time to the average cooling time calculated from the thermal energy plotted in panel 10(d) 
shows that $t_{\rm ff}/\bar{t}_{\rm cool} > 1$  for all the runs in this paper, even though $t_{\rm ff}/t_{\rm eddy}<1$ 
meaning that material will collapse vertically faster then gravitational driving will form clumps horizontally.
Thus, in agreement  with the earlier results of \citep{MSQP12} and \citet{SMQP12}, we expect a multiphase distribution.  
On the other hand, as Table~\ref{sumsim} shows, not all of these runs harbor strong outflows. Within the realm of our 
numerical setup,  horizontal driving always allows for $t_{\rm ff}/\bar{t}_{\rm cool} > 1$ and a multiphase medium, 
but this is not a sufficient condition for the existence of sustainable outflows.

\section{Conclusions}

Global outflows occur across a wide range of galaxy masses and redshifts, and 
observations suggest that they are most prominent in galaxies in which  the star-formation rate 
density exceeds $\dot{\Sigma}_\star \geqslant 0.1\,{\rm M}_\odot\,{\rm yr}^{-1}\,{\rm kpc}^{-2}$. 
Furthermore,  recent observations show that the ISM in such galaxies has internal velocity dispersions 
of $\approx 50-100\,{\rm km\,s^{-1}}$ \citep{SH09, West+09, Law+09, Green+10, Genzel+11, Swinbank+11}.
However, typically numerical simulations with supernova 
driving attain velocity dispersions of only $\approx 7-20\,{\rm km\,s^{-1}}.$ 
As current observations are unable to resolve the disk scale height or cleanly determine the orientation of
random velocities  with respect to the plane, it is thus possible that the 
large observed velocity dispersions arise not from stellar feedback but from large-scale gravitational instabilities 
in the disk. These will occur on scales $R\gg H$ and tap into macroscopic differences in the rotation rate and the 
gravitational potential, rather than nuclear processes, to drive large, anisotropic random motions.  

Here we have carried out a first exploration of the possible role of gravitational instabilities in driving
outflows from high surface-density disk galaxies. 
To focus on the effect of instabilities rather than the growth and maintenance of these modes, 
we drive turbulence to a level expected for a Toomre critical disk. 
Within this framework, we examine the role of two key parameters, namely the gas surface density, $\Sigma_{\rm g}$ 
and the epicyclic frequency, $\kappa$. Crucially, only the horizontal turbulence is driven, consistent with 
expectations for disk instabilities. 
The advantage of this parametrization, in which $\sigma_{\rm H} = \pi\,G\,\Sigma/\kappa$, 
is that we are able to control the velocity dispersion by simply adjusting the surface 
density of the medium. This allows us to probe a wide variety of disk conditions. 
Moreover as Figures~\ref{starform}, \ref{veldis} and \ref{tscales}
show, our approach also enables us to probe the variation of turbulent velocities, 
gas mass loss rates and the various timescales with the gas surface density and 
the epicyclic frequency. 

The key result arising from our study is that turbulence of the amplitude expected from gravitational 
instabilities can indeed drive galaxy outflows at a level $\dot{\Sigma}_{\rm g}/\dot{\Sigma}_{\star} \gtrsim 0.1$,
even in absence of additional energy input from supernovae. Our models show that global outflows of 
this kind are likely to be present in the highly compact and rapidly rotating disks that were common at 
high redshift, while they are more likely to be weak or completely absent in less compact and slowly 
rotating disks such as our Milky Way. Within the range of the parameter space of 
$\Sigma_{\rm g}$ and $\kappa^{-1}$ that we probed, 
$\dot{\Sigma}_{\rm g}\geqslant 0.02\,{\rm M}_\odot\,{\rm yr}^{-1}\,{\rm kpc}^{-2}$
 outflows occured in galaxies where the gas surface densities exceeded 
$100 \,{\rm M}_{\odot}\,{\rm pc^{-2}}$ with a lower bound of $\kappa^{-1}=20\,{\rm Myr}$. 
Interestingly, this threshold values is very close to the value of the critical star-formation rate 
densities  of $\dot{\Sigma}_{\star} > 0.1\,{\rm M}_\odot\,{\rm yr}^{-1}\,{\rm kpc}^{-2}$. 

Our results also indicate that outflows arise if the one-dimensional horizontal velocity dispersion 
exceeds a critical value of $\approx 35\,{\rm km\,s^{-1}}$.   
We find that the vertical velocity dispersion has a mean value of $5.7\,{\rm km\,s^{-1}}$, 
largely independent of the parameters of our simulation. Such low values are also obtained in ISM simulations 
where supernovae feedback is explicitly included \citep[e.g.,][]{SO12, KOK13}.
Furthermore, two-dimensional slices in the $x-z$ plane show their disks to be thin and quasi-steady
over many eddy-turnover times.  
In an additional test run, we lowered the temperature floor to 300 K for the S250K10 run and found 
that the mass outflow rate and the estimate of the mass-weighted $\sigma_{\rm z}$ are similar to the 
estimates from the fiducial S250K10 run (see Appendix for a comparison between the two runs). 
This implies that lowering the temperature floor to capture the dynamics of the cold dense medium is not 
crucial for outflow generation.

The occurrence of outflows in our study can be further explained from the thermodynamic properties of the 
ISM in these disks. As Figure~\ref{s500k10} shows, even if one starts with a constant temperature distribution, 
the ISM in these galaxies soon evolves into a multiphase distribution. Certain regions attain temperatures 
higher than the critical temperature beyond which the radiative cooling rate progressively 
decreases. Over the course of the evolution, this results in a runaway arising from inefficient cooling of these 
hot regions coupled with successive heating from turbulent driving. The combined action of these two effects 
leads to the motion of the gas outward through the simulation domain. In a nutshell, the above arguments 
suggest that in the absence of stellar feedback, outflows from high redshift galaxies can arise from
a turbulent heating instability due to the progressive  decline in the efficiency of radiative cooling beyond 
$T\approx 2.5\times 10^{5}\,{\rm K}$. 

As a caveat, we note that the results of this study {\it assume} that turbulence can be sustained at a steady 
level $\sigma_{\rm H} \approx \pi\,G\,\Sigma_{\rm g}/\kappa$ over many local dynamical times, via horizontal 
instabilities at scales $R\approx \pi\,G\,\Sigma_{\rm g}/\kappa^{2}$ between the disk 
thickness $(\sigma_{\rm z}/\sigma_{\rm H})^2R/2 $ and the Toomre wavelength 
$4 \pi R$. However, previous simulations have shown that without small-scale feedback, gravitationally unstable 
disks may have SFRs high enough to deplete the local gas within a few orbital times \citep[e.g.,][]{Hopkins+11, Agertz+13}. 
This would reduce the spatial scale ($\propto R$ if Q remains $\approx 1$) of gravitational instabilities and 
the corresponding rotational velocity differences that these instabilities can tap, leading to lower 
$\sigma_{\rm H} \approx\pi\,G\,\Sigma_{\rm g}/\kappa$. Thus, a state in which high amplitude turbulence is 
maintained over many orbital times may require significant accretion of gas from larger scales and/or suppression 
of local collapse by stellar feedback.  We also note here that turbulent driving as implemented in our 
simulations is insensitive to hot or cold gas in the sense that it is likely to drive both the high density cold 
gas and the low density hot gas in a similar fashion. However, in a realistic environment self-gravity could 
drive high density and low density gases very differently. The resulting effect on the generation of outflows 
in such systems merits a careful analysis.

An important limitation of our work is the neglect of cold dense phases of the ISM. An accurate modeling of 
the CNM requires the inclusion of cooling due to both fine structure atomic lines and CO rotational line emission. 
However, this would lead to the development of small, very dense structures that would not be spatially resolved 
in some of our present simulations. It would be interesting to include these low temperature processes in the 
cooling routines in a future study with the aim to understand its effect on the generation of galaxy outflows. 

An additional issue with the simulations reported here relates the number of cells by which the length scale 
$l_{\rm sg} \approx c_s^{2}/g(z)$,  corresponding to the gravity source term in the momentum equation, is resolved. 
Because the present  simulations do not follow 3D self-gravitating fragmentation at small scales, the usual 
Truelove criterion ($l_{\rm J} > 4\,dx$) for avoiding unphysical excitation of small-scale noise in AMR simulations 
does not directly apply. Here, we have  only vertical gravity associated with the integrated surface density of gas, 
which is not substantially altered  by grid-scale noise. However, we note that in our simulations we only just 
resolve the Jeans scale for low surface density disks. Thus, fully self-gravitating simulations that seek to 
address the issues considered in this paper would require substantially higher resolution than we have adopted 
here.

Our results also raise the question of how energy input from supernovae and gravitational instabilities might  
work in conjunction in rapidly star forming galaxies. The outflow rates obtained in our study should be 
thought of as lower limits; stellar processes may or may not lead to additional mass loss. 
Interestingly, the high-surface density disk simulations of \citet{WN07} showed that the inclusion 
of supernova feedback did not lead to any appreciable change in the morphology or in the density PDF 
(see Figs.~18 and 19 in their paper)  compared to the case where turbulence in the disk solely arose 
from gravitational instabilities. However, these simulations only spanned a region $0.32\,{\rm kpc}$ in the 
vertical direction which may not be sufficient to draw any immediate conclusions about the combined impact 
of stellar feedback and gravitational instabilities. Among other questions, it will be interesting to explore how 
turbulent velocities resulting from the combined effect of gravitational instabilities and supernova driving scale 
with disk properties. 
Correspondingly, it would be interesting to test how star formation rates differ when turbulence is 
driven by both supernovae and large-scale instabilities. Finally, magnetic fields are an important component 
of the ISM where they play a variety of roles ranging from controlling star formation, influencing turbulent 
mixing \citep{Sur+14, SPS14}, to the confinement and propagation of cosmic rays. How such fields will affect 
the conclusions of this study is an open question. These are some of the issues we intend to address in 
forthcoming papers. 

\acknowledgments

We would like to thank William Gray, Christopher Matzner, Prateek Sharma and Robert Thacker for helpful 
discussions and the anonymous referee for his/her comments. S.~S \& E.~S were supported by National 
Science Foundation grant AST11-03608 and NASA theory grants NNX09AD106 and NNX15AK82G. E.~S gratefully 
acknowledges the Simons Foundation for funding the workshop  {\em Galactic Winds: Beyond Phenomenology} 
which helped to inspire this work.  He also gratefully acknowledges Joanne Cohn, Eliot Quataert, and the UC Berkeley 
Theoretical Astronomy  Center, and Uro\v{s} Seljak and the Lawrence  Berkeley National Lab Cosmology group, for 
hosting him during the period when much of this work was carried out.  Part of this research was carried out during the 
visit of E.~S and E.~C.~O at the KITP in U.~C. Santa Barbara, which  is supported by the National Science 
Foundation under grant PHY-1125915. The work of E.~C.~O on this project was supported by the National 
Science Foundation under grant AST-1312006. The authors would like to thank the  Texas Advanced Computing 
Center (TACC) at The University of Texas at Austin (URL: http://www.tacc.utexas.edu),  and the Extreme 
Science and Engineering Discovery Environment (XSEDE) for providing HPC resources via grant 
TG-AST140004 that have contributed to the results reported within this paper. The FLASH code is developed 
in part by the DOE-supported Alliances Center for Astrophysical Thermonuclear Flashes (ASC) at the 
University of Chicago.

\bibliographystyle{apj}
\renewcommand{\leftmark}

\begin{appendix}

\section{Resolution Dependence}\label{Res}

\begin{table*}[h]
\begin{minipage}{170mm}
\begin{center}
\resizebox{0.6\textwidth}{!}{
\begin{tabular}{|c|c|c|c|c|c|} \hline \hline 
Simulation & $dz$ & $H$ & $\bar{\sigma}_{\rm H}^{\rm 1D}$ & $\bar{\sigma}_{\rm z}$ & $\dot{\Sigma_{\rm g}}$ \\
Resolution & [pc] & [pc] & $[\rm km\,s^{-1}]$ & $[\rm {km\,s^{-1}}]$ & $[{\rm M}_\odot\,{\rm yr}^{-1}\,{\rm kpc}^{-2}] $ \\ \hline \hline
$512^{2} \times 1024$ & 4.16 & 10 & $54$ & $5.5$ & $0.062$ \\ \hline
$256^{2} \times 512$   & 8.32 & 11 & $50$ & $5.5$ & $0.06$ \\ \hline 
$128^{2} \times 256$   & 16.6 & 14  & $51$ & $5.0$ & $0.057$ \\ \hline 
$64^{2} \times 128$     & 33.3 & $15.5$ & $47$ & $7.1$ & $0.054$ \\ \hline \hline 
\end{tabular}
}
\end{center}
\caption{Study of resolution dependence for run S250K10. }
\label{resdep}
\end{minipage}
\end{table*}

\begin{figure}[t]
\centering
\includegraphics[width=0.8\columnwidth]{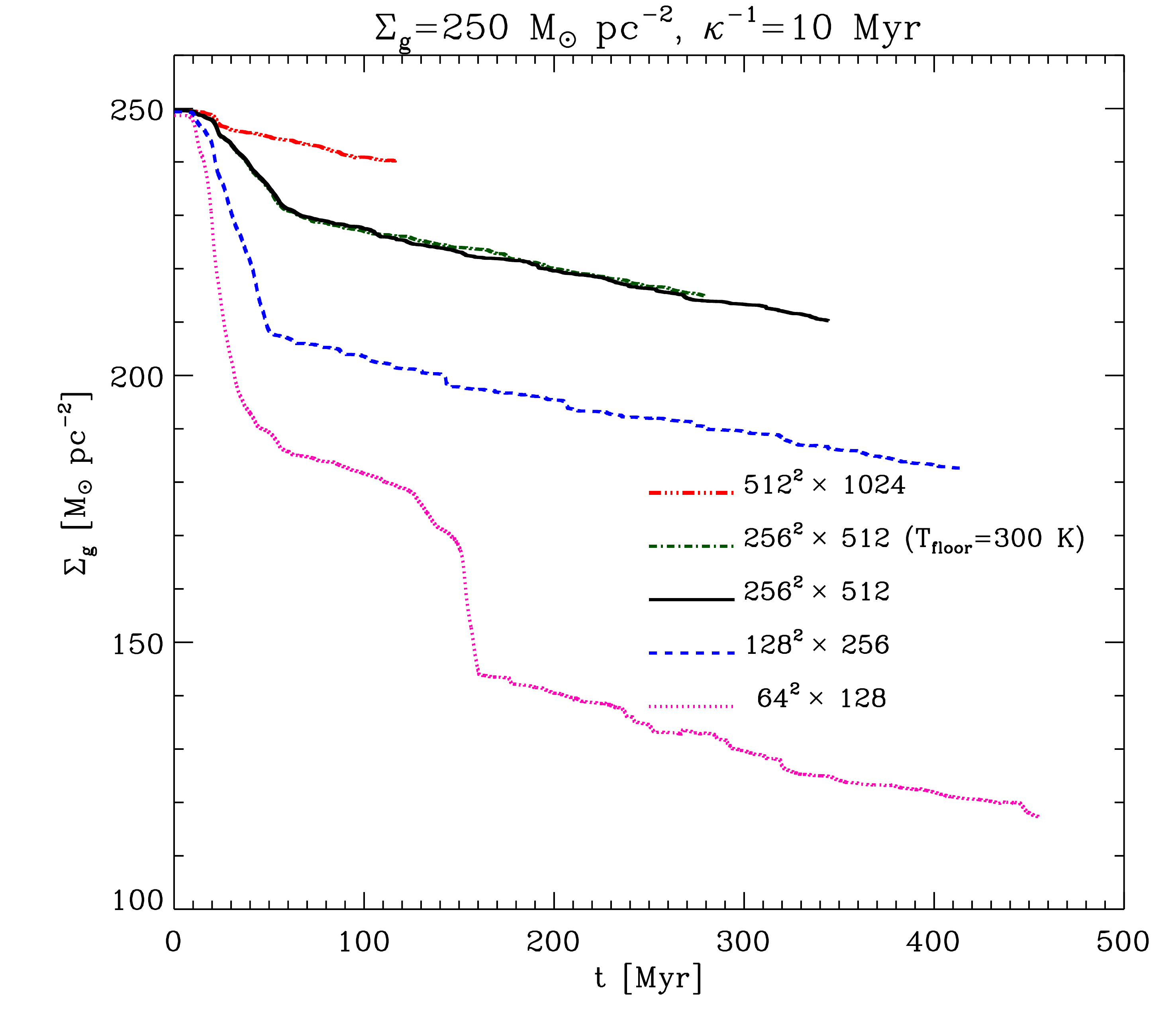} 
\caption{(color online). Time evolution of the gas surface density at four different resolutions for initial 
$\Sigma_{\rm g} = 250\,{\rm M_\odot\,pc^{-2}}$, and $\kappa^{-1} = 10\,{\rm Myr}$. 
Note that the gas mass loss rate nears convergence from a resolution of $256^{2}\times 512,$ with $dz=8.32$ 
pc as compared to a vertical scale-height of 11 pc. It is also clear that the run with a temperature floor of 
$300$ K has the same gas mass loss rate as our fiducial S250K10 run. 
}
\label{mloss_resdep}
\end{figure}

To explore the impact of resolution effects, we performed a resolution study for our fiducial high surface density 
case: S250K10, with $\Sigma_{\rm g}$ = 250 ${\rm M}_\odot$ pc$^{-2}$ and $\kappa^{-1}$ = 10 Myr.  This  
involved a single high-resolution $512^{2}\times 1024$ run, with $dz$ half that of our standard, $256^{2}\times 512$ 
run, and two low-resolution runs with $128^{2} \times 256$ and  $64^{2}\times 128$ cells, and $dz$ twice and four 
times as large as in our standard runs, respectively. In Table~\ref{resdep}, we show a comparison of the computed 
scale-height, one-dimensional horizontal and vertical velocity dispersions, and gas mass loss rate obtained in these 
four simulations. We also show the time evolution of $\Sigma_{\rm g}$ for each of the runs in Figure~\ref{mloss_resdep}. 

In all four runs, our turbulent driving results in a similar 1D horizontal vertical dispersion $\approx 50$ km/s.
Table  \ref{resdep} also shows that the scale-heights, vertical velocity dispersion, and
gas mass loss rate are very similar between the standard, $256^{2}\times 512$  run and the high-resolution, 
$512^{2}\times 1024$ run in which the scale-heights were resolved by $\approx 10$ and $20$ cells, respectively.  
For the high-resolution run, the gas mass loss rate is computed from $84$ Myr to the end of the simulation. 
However, as evident from Figure \ref{mloss_resdep}, the above two runs differ in their evolution in the initial 
transient phase. While the $256^{2}\times 512$ loses about $8.7\%$ of its initial mass during this phase, the 
$512^{2}\times 1024$ run loses much less mass during its initial rearrangements. Therefore, barring the estimates 
during the transient phase, the gas mass loss rate is similar in both the runs. This gives us confidence that 
the results reported in the paper are not strongly influenced by resolution effects, and that even larger and 
more expensive simulations are not required to reach reliable conclusions. 
Moreover, by comparing the 
green dash-dotted line with the black line, it appears that lowering the temperature floor to $300$ K does not 
result in any change in the mass outflow rate. Thus, our conclusions regarding the mass outflow rates are 
not likely to be affected by modeling the low temperature ISM.

Moving to the lower resolution $128^{2} \times 256$ run, we find that the estimates of the horizontal velocity dispersion, 
vertical  velocity dispersion, and gas mass loss rate is very close to those of the two higher resolution runs. However, 
the estimate of the scale height falls below the resolution limit and the gas mass loss rate in the transient phase
is more than the higher resolution runs. For the lowest resolution $64^{2} \times 128$ run, as Figure~\ref{mloss_resdep} 
shows, the gas mass loss rate undergoes an extremely sharp decline for about $100$ Myr followed by a steep 
decline from about $120-160$ Myr. During this phase, the temperature also drops sharply from 
$1.13\times 10^{6}\,{\rm K}$ to $7.8\times 10^{5}\,{\rm K}$, after which the system reaches a steady state. 
The mass loss rate shown in the table is computed from $350$ Myr onwards.  
In this case, the computed value of the scale height is well below the resolution limit. 

\begin{figure}[t]
\centering
\includegraphics[width=1.02\columnwidth]{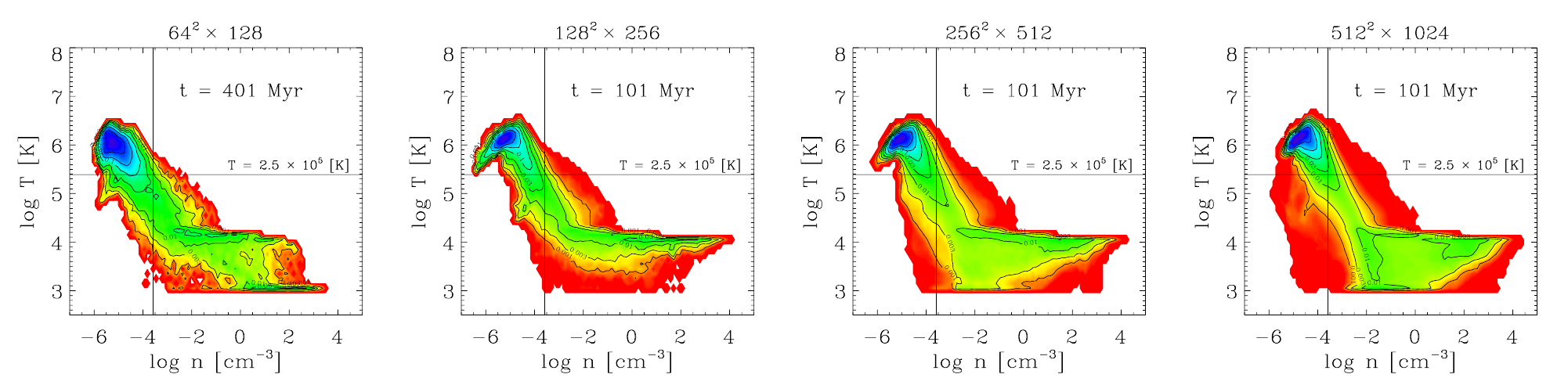} 
\caption{(color online). Volume-weighted phase diagrams of the temperature and the number 
density of all the gas at four different resolutions for the S250K10 run. The vertical and the 
horizontal lines have the same meaning as in Figure~\ref{phaseplots}. At both $256^{2}\times 512$
and $512^{2}\times 1024$ resolutions, the phase plots show an accumulation of low density gas 
at $1000$ K as compared to the $128^{2}\times 256$ run. 
}
\label{phase_resdep}
\end{figure}

In Figure~\ref{phase_resdep}, we show the volume-weighted $T-n$ phase plots over the whole simulation domain 
at different resolutions.  The phase  plots are shown at $101$ Myr,
except for the lowest resolution run, which takes much longer to reach a steady-state, and whose phase plot
 is shown at $401$ Myr. From these diagrams it appears that the phase plots are similar for the 
 $256^{2}\times 512$ and the $512^{2}\times 1024$ resolutions. However compared to the $128^{2}\times 256$ 
run, both of these runs show an accumulation of low density gas at $1000$ K. The $128^{2}\times 256$ 
resolution run also shows some gas at $T = 2.5\times 10^{5}\,{\rm K}$ being dragged along with the hot outflowing gas. 
On the other hand, the phase plot at $64^{2}\times 128$ is very different from the other three, the most significant 
being the accumulation of high density gas at the temperature floor.

Thus, while the results presented here are not likely to suffer from resolution effects, they appear to be close to the lowest 
resolution allowable to achieve reliable results, even in the absence of additional physical processes such as stellar feedback, 
molecular chemistry, cosmic ray heating, and magnetohydrodynamic effects. This means that  cosmological simulations that 
are unable to achieve $\approx 10$ pc resolutions  will not be able to properly handle the evolution of the ISM in high-surface 
density galaxies similar to the S250K10 case, and may instead either over or under-represent gas mass loss depending on the 
particulars of the numerical method being used.  Furthermore, even higher resolutions are likely to be required to model 
outflows from higher surface density disks with smaller vertical scale heights, even in the case without supernovae.

\end{appendix}

\end{document}